\documentclass{aa} 
\usepackage{graphicx} 
\usepackage[varg]{txfonts}
\usepackage{xcolor}
\usepackage[normalem]{ulem}
\usepackage{placeins}
\usepackage{hyperref}
\usepackage{siunitx}
\usepackage{float}
\newsavebox{\dummybox}
\newcommand{\invisobject}[1]{\sbox{\dummybox}{\object{#1}}}

\begin{document}
\makeatletter
\let\linenumbers\relax
\let\endlinenumbers\relax
\makeatother

   \title{Multiband analysis of the O’Connell effect in 14 eclipsing binaries}

   \subtitle{}

   \author{D. Flores Cabrera\inst{1,2}
        \and M. Catelan\inst{1,2,3}
        \and A. Papageorgiou\inst{4}
        \and A. Clocchiatti\inst{1,2}
        }

   \institute{Instituto de Astrofísica, Pontificia Universidad Católica de Chile, Av. Vicuña Mackenna 4860, 782-0436 Macul, Santiago, Chile
    \email{ddflores@uc.cl, mcatelan@uc.cl}
    \and Millennium Institute of Astrophysics, Nuncio Monse\~{n}or Sotero Sanz 100, Of. 104, Providencia, Santiago, Chile
    \and Centro de Astroingenier{\'{\i}}a, Pontificia Universidad Cat{\'{o}}lica de Chile, Av. Vicu\~{n}a Mackenna 4860, 7820436 Macul, Santiago, Chile
    \and Department of Physics, University of Patras, 26500, Patra, Greece}

   \date{Received September 30, 20XX}

  \abstract
   {The O'Connell effect is a phenomenon in eclipsing binary (EB) systems that consists of unequal maxima in a light curve when out of eclipse. Despite being known for decades and with several theories proposed over the years, this effect is still not fully understood.}
   {Our goal is to find different O'Connell effect properties using a multiband approach, compare them with each other, and find correlations between these properties and the physical parameters of the systems.}
   {We present the analysis of 14 new EBs that show the O'Connell effect using multiband data extracted from the Asteroid Terrestrial-impact Last Alert System (ATLAS) and \textit{Zwicky} Transient Facility (ZTF) all-sky surveys. We measured the difference in maximum amplitudes ($\delta m$) alongside different light curve features in different passbands via a new modeling process that uses Gaussian fits. We created a brand-new phenomenological model for O'Connell effect systems whose analysis had previously been hampered by overfitting.}
   {Although the magnitude of the O'Connell effect seems to be more pronounced at shorter effective wavelengths, supporting the idea that cool starspots cause the effect, a conclusive correlation is not found. On the other hand, we do find strong correlations between the magnitude of the effect and both the temperature and the period, both of which are inconsistent with a previous study. We also find that in systems that show both positive and negative O'Connell effects, there are different correlations with the aforementioned parameters.}
   {We conclude that, even though starspots may be one cause of the O'Connell effect, it is likely a multipronged phenomenon; for instance, the physical interaction between the components of close binary stars may be another important factor.}


   \maketitle

\section{Introduction}\label{sec:intro}
Binary stars have played a key role in stellar astrophysics as they provide a means to accurately measure the physical parameters of stars, including their masses. In the case of eclipsing binaries (EBs), the light curves (LCs) embed information about the component stars' radii, temperatures, and luminosities. Among the least understood characteristics of EB LCs is the fact that the out-of-eclipse maxima are different. This asymmetry is known as the O'Connell effect \citep{OConnell1951}. Although it affects a relatively small fraction of EBs \citep[of order 7\%, according to][]{Knote_2022}, it complicates the measurement of their physical parameters, leading to errors in estimated stellar radii, luminosities, and orbital inclinations. Furthermore, the effect may mimic the presence of a third body, further complicating interpretation.

\citet{OConnell1951} found, based primarily on visual and photographic data, predominantly brighter first maxima. \citet*{1984ApJS...55..571D} revisited and expanded his work using $UBV$ photoelectric photometry, arriving at somewhat different conclusions. In particular, they observed a more even distribution of asymmetries and found that the brighter maximum was often redder. Their study suggested that multiple mechanisms, including magnetic activity, circumstellar material, and tidal distortions, can contribute to the effect, thus highlighting the need for further investigation.

Several interpretations have been proposed to explain the O'Connell effect. They include the presence of cool stellar spots due to the magnetic activity of the component(s) \citep{1960AJ.....65..358B}, hot spots as the result of gas stream impact due to mass transfer \citep{1995AcASn..36...37Z, 1997MNRAS.291..749H, 2011AN....332..607P}, and clouds of circumstellar material \citep[dust and gas;][]{Liu_2003}. 
In particular, cool spots have been demonstrated as a viable explanation by means of Doppler imaging techniques \citep[e.g.,][]{Hendry_2000,Senavci_2011,Niu_2022}. The other two scenarios have gained popularity recently. The magnetic activity interpretation applies more to late-type systems due to the presence of deep convective zones and rapid rotation. For earlier-type contact systems (A-F spectral types), on the other hand, the LC asymmetries are more probably due to mass transfer from the more massive star: as the primary star transfers material through the first Lagrangian point to its companion, the impact generates a region of enhanced brightness on the receiving star. When this hot spot rotates into view during the orbital cycle, it causes one of the LC maxima to appear brighter than the other, thus producing the O'Connell effect \citep{2011AN....332..607P}.
 
Several models have been used for the LC phenomenological modeling of EBs \citep[e.g.,][]{2008ApJ...687..542P, 2015OAP....28..181T,2015A&A...584A...8M} and/or to search for contact systems presenting the O'Connell effect  \citep{2009SASS...28..107W, 2019ComAC...6....4J, 2014CoSka..43..470P, 2017EPJWC.15203010P,Knote2022}. One of the primary quantitative indicators of the presence of the O'Connell effect is $\Delta m_{\text{max}}$, defined as the difference in magnitude between the two maxima of the LC. Using \textsc{Hipparcos} contact binaries presenting the O'Connell effect, \cite{2011AN....332..607P} showed that, for late-type systems, the signs of $\Delta m_{\text{max}}$ at quadrature phases are random, while for the earlier systems there is a clear tendency for the first quadrature maximum to be brighter than the second. 

Although $\Delta m_{\text{max}}$ can give us some information about how strong the effect is in different systems, other useful indicators have also been defined. In particular, \cite{1999PhDT........38M} introduced the O'Connell effect ratio (OER) and light curve asymmetry (LCA) metrics, both of which provide further insight into the asymmetries of the LCs. OER is defined as the ratio of the area under the curve for phases $\phi=0-0.5$ and $\phi=0.5-1$. This means that a LC with an $\text{OER}=1$ has the same amount of flux in both halves, while an $\text{OER}\neq 1$ means that one of the two halves has more flux than the other, indicating an asymmetry. LCA quantifies deviations from symmetry in EB LCs by directly assessing the point-wise asymmetry between corresponding phase bins. Unlike the OER, which measures total flux differences, LCA provides a more localized evaluation of asymmetry. In like vein, LCA values greater than $0$ indicate that the LC is asymmetric, whereas $\text{LCA}=0$ means that the LC is highly symmetric.

This work presents a multiband LC analysis of the O'Connell effect in 14 EBs using photometric data from the \textit{Zwicky} Transient Facility \citep[ZTF;][]{Bellm_2019,Masci_2019} and the Asteroid Terrestrial-impact Last Alert System \citep[ATLAS;][]{2018PASP..130f4505T}. We also present a new way to model LCs with strong asymmetries, which is different from the commonly used Fourier series fit, and provide the OER, LCA, and $\Delta m_{\rm max}$ parameters for each of these systems in four different passbands. 

In Sect.~\ref{sec:obs} we present the data used in our study. In Sect.~\ref{sec:selection} we describe our procedure for identifying O'Connell EBs therefrom. Our new phenomenological model for these systems is described in Sect.~\ref{sec:Model}, and the results of its application to our multiband LCs are presented in Sect.~\ref{sec:results}. Finally, a discussion of our results and conclusions can be found in Sect.~\ref{sec:conclusions}.

\section{Observations}\label{sec:obs}
The data used in this project came from two main sources, namely the ZTF and ATLAS surveys. 

\subsection{ZTF}

ZTF is a Samuel Oschin 48-inch Schmidt telescope located at Palomar Observatory with a 47~deg$^{2}$ field of view \citep[FOV;][]{Bellm_2019,Masci_2019}. This large FOV ensures that ZTF can scan the entire northern sky every night. It can observe 3760~deg$^{2}$ per hour, reaching a 5$\sigma$ detection limit of 20.5~mag in $r$ with a 30\,s exposure. Its primary science aims are the physics of transient objects, stellar variability, and Solar System science \citep{Graham_2019}. ZTF photometry for public surveys is obtained in two observation modes in the $g$, $r$, and $i$ bands, namely three-night cadence for the entire visible northern sky from Palomar and one-night cadence for the Galactic plane (Galactic latitude $|b| \leq 7^{\circ}$). In our analysis, the dataset that was used corresponds to ZTF Data Release 15 (DR15). Perfectly clean extractions ($\texttt{catflag} = 0$, $\texttt{|sharp|} < 0.25$, and $\texttt{mag} < \texttt{limitmag} - 1.0$) were used in both $g$ and $r$ bandpasses at every epoch. Since ZTF's DR15 did not include $i$-band data for many of our sources, we did not include this band in our analysis. DR15 contains on average $\sim 637$ epochs of observations in $g$ and $\sim 850$ in $r$ per star in our sample, with observations covering the time span between March 2018 and November 2022.

\subsection{ATLAS}

ATLAS is a survey designed to detect small asteroids (10–140~m) on their ``final plunge'' toward impact with Earth \citep{2018PASP..130f4505T}. It observes about 13000 deg$^2$ of sky at least four times per night, which also makes the resulting dataset very useful for the discovery of transient events and the follow-up of variable stars. ATLAS uses Wright Schmidt telescopes with a 4.5~deg$^2$ FOV each, and each ATLAS telescope takes four 30\,s exposures of 200–250 target fields per night. Although ATLAS is tuned for investigating near-Earth asteroids, it collects the LCs of transients and variable stars down to a magnitude limit of $r$ $\sim$ 18 in the $c$ ($\sim g+r$) and $o$ ($\sim r+i$) bands, for which approximate color transformations linking them to the Panoramic Survey Telescope and Rapid Response System (Pan-STARRS) $g$, $r$, and $i$ bands \citep{Magnier2020} were provided by \citet{2018PASP..130f4505T}. 

ATLAS observations of all our objects were taken between July 2015 and October 2020. 
The ATLAS photometry we incorporated in our analysis was extracted directly from the ASCII files in the ATLAS main storage computer.
Finding our objects of interest required going through the ATLAS structured query language (SQL) database, which organizes the exposures based on multiple keywords.
A coordinate-based search for the stars previously selected in the ZTF survey was performed in the database to locate the exposures that include the objects of interest. Knowledge of the exposures allowed us to locate the final photometry catalogs of those images (the ``.dph'' files).
We developed scripts to scan these catalogs for detections matching the object coordinates within one ATLAS pixel, and the final photometry of the matching objects was extracted together with the exposure dates.

\invisobject{ZTF J020930.46+513905.4}
\invisobject{ZTF J014355.46+354402.7}
\invisobject{ZTF J054353.09+014433.7}
\invisobject{ZTF J093054.62+745619.7}
\invisobject{ZTF J091456.62+394615.5}
\invisobject{ZTF J011757.02+305050.8}
\invisobject{ZTF J164516.11+321905.2}
\invisobject{ZTF J005348.21+465518.6}
\invisobject{ZTF J181310.67+761840.2}
\invisobject{ZTF J155632.27+654934.9}
\invisobject{ZTF J201843.21+540656.4}
\invisobject{ZTF J200516.81+520253.7}
\invisobject{ZTF J014744.51+450636.8}
\invisobject{ZTF J182814.00+551351.2}

\begin{table*}[ht]
\caption{O'Connell effect candidates detected in this work.}
\label{tab:ztflist}
\begin{tabular}{lccccccccc}
\hline\hline
ID & R.A. (J2000) & Dec. (J2000) & \multicolumn{4}{c}{Mean Magnitude} & $P$ & Epoch\tablefootmark{a} & $T_{\rm eff}$\\ \cline{4-7}
 & h m s & $\degr$ $'$ $''$ & $g$ & $c$ & $r$ & $o$ & (days) & MJD & (K) \\ 
\hline
ZTF17aaadztm & 02 09 30.462 & +51 39 05.30 & 17.509 & 16.976 & 16.479 & 16.378 & 0.251132 & 58716.442 & 4617 $\pm$ 159\\
ZTF17aaaesno & 01 43 55.461 & +35 44 02.58 & 15.309 & 14.967 & 14.601 & 14.520 & 0.269418 & 58836.098 & 5033 $\pm$ 268\\
ZTF17aabvxoo & 05 43 53.094 & +01 44 33.77 & 16.733 & 16.204 & 15.739 & 15.606 & 0.287529 & 58766.522 & 4702 $\pm$ 218\\
ZTF18aaacuaj & 09 30 54.619 & +74 56 19.62 & 17.057 & 16.842 & 16.508 & 16.512 & 0.263626 & 58894.235 & 5541 $\pm$ 174\\
ZTF18aaaclzz & 09 14 56.623 & +39 46 15.58 & 16.655 & 16.241 & 15.854 & 15.753 & 0.226229 & 58877.311 & 4821 $\pm$ 145\\
ZTF18aaakvwc & 01 17 57.013 & +30 50 50.70 & 16.405 & 15.943 & 15.457 & 15.332 & 0.246259 & 58724.393 & 4599 $\pm$ 154\\
ZTF18aabexej & 16 45 16.125 & +32 19 05.07 & 16.986 & 16.707 & 16.375 & 16.340 & 0.293578 & 58586.382 & 5313 $\pm$ 180\\
ZTF18aabfljv & 00 53 48.221 & +46 55 18.53 & 17.145 & 16.841 & 16.519 & 16.447 & 0.272984 & 59118.242 & 5262 $\pm$ 181\\
ZTF18abaaway & 18 13 10.667 & +76 18 40.16 & 15.439 & 15.202 & 14.858 & 14.822 & 0.274787 & 58557.378 & 5426 $\pm$ 154\\
ZTF18aagrehv & 15 56 32.274 & +65 49 34.89 & 16.329 & 16.029 & 15.733 & 15.728 & 0.281808 & 58294.242 & 5310 $\pm$ 139\\
ZTF17aaawfrc & 20 18 43.208 & +54 06 56.25 & 17.277 & 16.773 & 16.346 & 16.180 & 0.263616 & 58678.466 & 5021 $\pm$ 163\\
ZTF17aabumbt & 20 05 16.821 & +52 02 53.63 & 15.586 & 15.279 & 14.945 & 14.887 & 0.297028 & 58441.192 & 5294 $\pm$ 215\\
ZTF18aabfkyu & 01 47 44.508 & +45 06 36.68 & 16.064 & 15.661 & 15.190 & 15.091 & 0.250981 & 59115.356 & 4816 $\pm$ 68\\
ZTF18aapmolr & 18 28 14.012 & +55 13 51.13 & 15.607 & 15.322 & 14.968 & 14.914 & 0.303991 & 58268.429 & 5198 $\pm$ 157\\
ZTF18abjhbjy & 20 39 53.984 & +22 39 10.95 & 17.627 & 17.151 & 16.833 & 16.698 & 0.300797 & 58257.485 & 4850 $\pm$ 155\\
ZTF18abumdef & 22 46 37.991 & +77 52 51.83 & 16.198 & 15.734 & 15.503 & 15.443 & 0.341751 & 59118.256 & 5238 $\pm$ 168\\
\hline
\end{tabular}
\tablefoot{
\tablefoottext{a}{Epoch of a minimum in brightness.}
}
\end{table*}

The photometry in ATLAS .dph files was done with a version of DoPhot \citep{1993PASP..105.1342S} that was refurbished by \citet{2012AJ....143...70A}, with further improvements by John Tonry (2016, private communication).
The .dph files are the final stage of the ATLAS reduction pipeline for non-moving objects; it builds upon previous stages that provide astrometric and photometric
solutions based on several different catalogs \citep{2018ApJ...867..105T} and photometric zero points that vary across the large charge-coupled device (CCD) field and include correction for nonuniform cloud cover.
More information on ATLAS photometry is given by \citet{2018AJ....156..241H}. 

\section{Candidate selection}\label{sec:selection}
Our initial search sample consisted of 640 objects (primarily on the Galactic plane) with ZTF $g$- and $r$-band LCs, classified as detached or semidetached EBs by the ALeRCE broker \citep{Forster2021,2021AJ....161..141S}. These objects, retrieved via direct database queries between April and June 2020, were required to have a minimum of 150 combined $g$+$r$ ZTF detections, as reported by ALeRCE, to ensure well-sampled LCs. All the objects were reclassified as contact binaries in the ZTF Catalog of Periodic Variable Stars \citep{chen2020zwicky}. The O'Connell effect is strongly associated with close binary interactions, particularly in contact systems where mechanisms like mass transfer, gas streams, and tidal distortions dominate. Accordingly, this sample provided a suitable initial basis for identifying candidates exhibiting LC asymmetries indicative of the O'Connell effect.

\begin{figure*}[t]
\centering
 \includegraphics[width=0.9\textwidth]{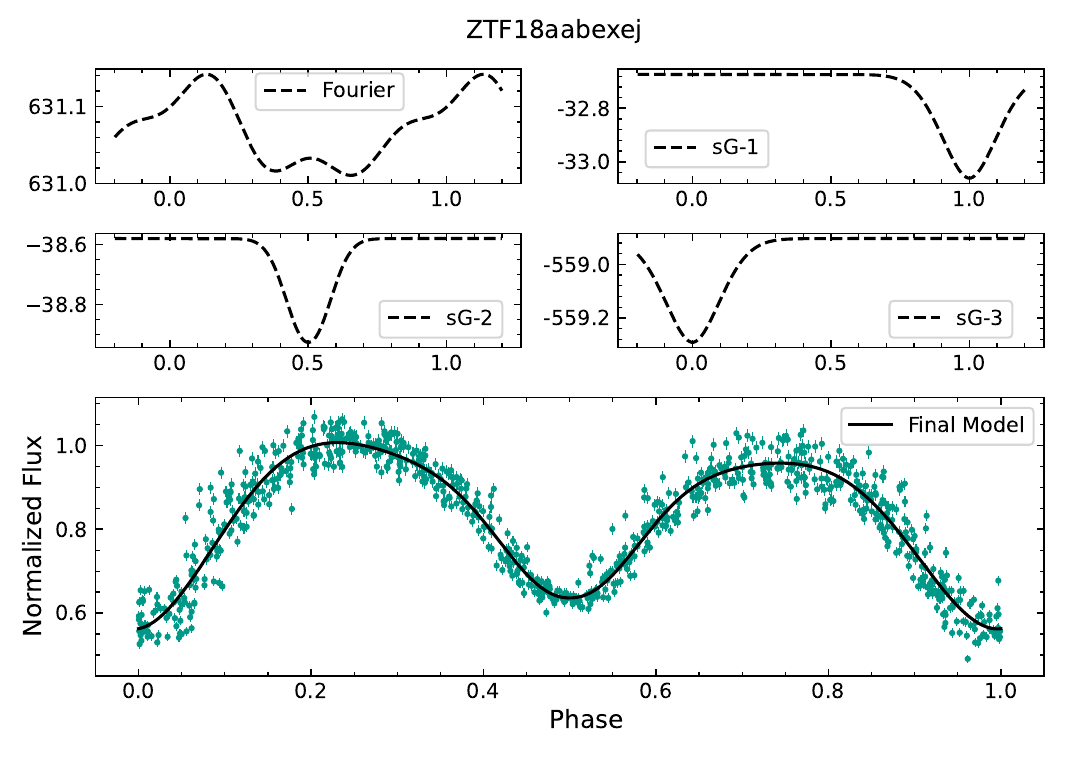}
 \caption{Phenomenological modeling of the O'Connell EB candidate ZTF18aabexej. Top panels: Four components of our model, as described in the text (Sect.~\ref{sec:Model}). Bottom panel: final model fit to ZTF $g$-band data.}
 \label{fig:mod}
\end{figure*}

We then performed an automated search for the O’Connell effect on all 640 candidates using the code provided by \cite{2014CoSka..43..470P, 2017EPJWC.15203010P}, which computes $\Delta m_{\rm max}$ values for each star. At this stage, the periods reported by \citet{chen2020zwicky} were used. Candidates were chosen that satisfied the criterion $\Delta m_{\rm max} > 0.05$~mag in ZTF's $r$ band; visual inspection of the LCs revealed that this cutoff is sufficient to reliably identify the presence of the O'Connell effect in the ZTF light curves. We visually inspected all the remaining ZTF DR15 LCs, ending up with a sample of 16 EBs that show clear evidence of the O'Connell effect. These stars are listed in Table~\ref{tab:ztflist}. Temperatures were found by cross-matching this sample with Transiting Exoplanet Survey Satellite (\textit{TESS}) Input Catalog Version 8 \citep{2019AJ....158..138S}. Out of these 16 stars, 14 with well-defined photometry available in ATLAS were retained for further analysis.

Finally, in order to refine the periods of the sources, we used the photometry in ZTF and ATLAS to compute a multiband Lomb-Scargle periodogram \citep{VanderPlas_2015}. This was accomplished using the \texttt{gatspy} package \citep{jake_vanderplas_2016_47887} in python. The periods listed in Table~\ref{tab:ztflist} are the result of this analysis. In all cases, they are very similar to the period values previously reported by \cite{chen2020zwicky}.

{\fontsize{5pt}{7pt}\selectfont\setlength{\tabcolsep}{3pt}
\renewcommand\arraystretch{0.9}
\begin{table*}[ht]
\caption{O'Connell effect quantifiers.}
\label{tab:t2}
\begin{tabular}{
  l
  c
  S[table-format=1.3]
  S[table-format=3.0(1)e-4]
  S[table-format=3.0(1)e-4]
  S[table-format=3.0(1)e-4]
  S[table-format=3.0(1)e-4]
}
\hline\hline
ID & Filter & {Mean flux} & {$\Delta I_{\text{sG}}$} & {$\Delta I_{\text{poly}}$} & {OER} & {LCA} \\
 & & {(normalized)} & & & \\
\hline
ZTF17aaadztm & $r$ & 0.849 & 0.0747(00028) & 0.0704(00034) & 0.8651(00091) & 0.0352(00003) \\
 & $g$ & 0.807 & 0.0843(00055) & 0.0859(00058) & 0.8152(00098) & 0.0468(00003) \\
 & $c$ & 0.834 & 0.0725(00152) & 0.1337(00175) & 0.7880(00080) & 0.0614(00003) \\
 & $o$ & 0.842 & 0.0747(00056) & 0.0850(00037) & 0.8183(00091) & 0.0463(00003) \\
ZTF17aaaesno & $r$ & 0.837 & 0.0183(00022) & 0.0224(00035) & 0.9503(00100) & 0.0286(00003) \\
 & $g$ & 0.825 & 0.0374(00023) & 0.0370(00035) & 0.9275(00095) & 0.0268(00002) \\
 & $c$ & 0.838 & 0.0232(00042) & 0.0573(00073) & 0.9236(00095) & 0.0248(00002) \\
 & $o$ & 0.848 & 0.0237(00029) & 0.0057(00034) & 0.9598(00098) & 0.0216(00002) \\
ZTF17aabvxoo & $r$ & 0.845 & 0.0481(00028) & 0.0438(00038) & 0.9045(00096) & 0.0253(00002) \\
 & $g$ & 0.821 & 0.0659(00050) & 0.0602(00055) & 0.8891(00093) & 0.0325(00002) \\
 & $c$ & 0.871 & 0.0545(00073) & 0.0487(00054) & 0.9045(00102) & 0.0250(00002) \\
 & $o$ & 0.859 & 0.0428(00040) & 0.0277(00031) & 0.9197(00096) & 0.0212(00002) \\
ZTF18aaacuaj & $r$ & 0.861 & -0.0109(00026) & -0.0141(00033) & 1.0452(00132) & 0.0125(00001) \\
 & $g$ & 0.851 & -0.0190(00037) & -0.0240(00037) & 1.0509(00129) & 0.0147(00001) \\
 & $c$ & 0.850 & -0.0588(00195) & -0.0592(00189) & 1.1204(00123) & 0.0461(00003) \\
 & $o$ & 0.879 & -0.0223(00100) & -0.0211(00065) & 1.0366(00119) & 0.0432(00003) \\
ZTF18aaaclzz & $r$ & 0.826 & 0.0382(00029) & 0.0394(00074) & 0.9063(00121) & 0.0676(00007) \\
 & $g$ & 0.818 & 0.0529(00039) & 0.0555(00084) & 0.8433(00110) & 0.0725(00009) \\
 & $c$ & 0.808 & 0.0911(00073) & 0.0758(00069) & 0.8546(00095) & 0.0440(00003) \\
 & $o$ & 0.840 & 0.0457(00041) & 0.0278(00038) & 0.9350(00101) & 0.0228(00001) \\
ZTF18aaakvwc & $r$ & 0.846 & 0.0545(00025) & 0.0529(00050) & 0.8996(00095) & 0.0267(00002) \\
 & $g$ & 0.818 & 0.0821(00035) & 0.0700(00064) & 0.8648(00089) & 0.0475(00003) \\
 & $c$ & 0.836 & 0.0515(00076) & 0.0652(00096) & 0.8142(00095) & 0.0688(00008) \\
 & $o$ & 0.853 & 0.0390(00033) & 0.0205(00039) & 0.8694(00099) & 0.0429(00004) \\
ZTF18aabexej & $r$ & 0.838 & -0.0385(00026) & -0.0398(00035) & 1.1109(00134) & 0.0240(00002) \\
 & $g$ & 0.820 & -0.0480(00033) & -0.0452(00048) & 1.1262(00136) & 0.0276(00002) \\
 & $c$ & 0.832 & -0.0423(00085) & -0.0486(00104) & 1.1646(00135) & 0.0417(00004) \\
 & $o$ & 0.845 & -0.0380(00046) & -0.0316(00042) & 1.0835(00133) & 0.0267(00002) \\
ZTF18aabfljv & $r$ & 0.832 & -0.0329(00032) & -0.0299(00037) & 1.0734(00149) & 0.0136(00001) \\
 & $g$ & 0.832 & -0.0532(00051) & -0.0497(00042) & 1.1646(00156) & 0.0362(00003) \\
 & $c$ & 0.837 & -0.0527(00106) & -0.0103(00087) & 1.0678(00136) & 0.0500(00003) \\
 & $o$ & 0.855 & -0.0181(00057) & -0.0370(00042) & 1.0690(00144) & 0.0313(00002) \\
ZTF18abaaway & $r$ & 0.869 & -0.0310(00026) & -0.0307(00028) & 1.1096(00133) & 0.0159(00001) \\
 & $g$ & 0.860 & -0.0429(00028) & -0.0411(00035) & 1.1344(00138) & 0.0208(00001) \\
 & $c$ & 0.853 & -0.0201(00081) & -0.0212(00113) & 1.0696(00122) & 0.0106(00001) \\
 & $o$ & 0.875 & -0.0141(00052) & -0.0531(00039) & 1.1240(00147) & 0.0211(00002) \\
ZTF18aagrehv & $r$ & 0.883 & -0.0477(00022) & -0.0495(00030) & 1.1951(00141) & 0.0292(00002) \\
 & $g$ & 0.865 & -0.0552(00029) & -0.0591(00038) & 1.2377(00142) & 0.0357(00003) \\
 & $c$ & 0.894 & -0.0882(00066) & -0.0915(00111) & 1.4680(00185) & 0.0478(00002) \\
 & $o$ & 0.868 & -0.0643(00043) & -0.0980(00033) & 1.3560(00169) & 0.0379(00002) \\
ZTF17aaawfrc & $r$ & 0.849 & -0.0345(00032) & -0.0313(00035) & 1.0785(00115) & 0.0168(00001) \\
 & $g$ & 0.851 & -0.0324(00059) & -0.0273(00046) & 1.0623(00121) & 0.0178(00001) \\
 & $c$ & 0.873 & -0.0455(00096) & -0.0969(00068) & 1.1513(00127) & 0.0480(00003) \\
 & $o$ & 0.865 & -0.0416(00052) & -0.0550(00038) & 1.0972(00115) & 0.0319(00002) \\
ZTF17aabumbt & $r$ & 0.829 & 0.0379(00024) & 0.0378(00039) & 0.9247(00102) & 0.0299(00003) \\
 & $g$ & 0.803 & 0.0426(00037) & 0.0323(00055) & 0.9249(00101) & 0.0343(00003) \\
 & $c$ & 0.806 & 0.0368(00047) & 0.0181(00061) & 0.9653(00107) & 0.0133(00001) \\
 & $o$ & 0.835 & 0.0140(00028) & 0.0023(00026) & 0.9835(00107) & 0.0071(00001) \\
ZTF18aabfkyu & $r$ & 0.854 & 0.0075(00026) & 0.0146(00055) & 0.9892(00118) & 0.0207(00002) \\
 & $g$ & 0.849 & 0.0129(00032) & 0.0033(00067) & 0.9864(00118) & 0.0146(00001) \\
 & $c$ & 0.845 & 0.0052(00056) & 0.0132(00096) & 1.0567(00143) & 0.0829(00005) \\
 & $o$ & 0.877 & 0.0045(00033) & 0.0085(00033) & 1.9973(00123) & 0.0598(00004) \\
ZTF18aapmolr & $r$ & 0.871 & -0.0458(00016) & -0.0383(00037) & 1.1328(00124) & 0.0317(00003) \\
 & $g$ & 0.851 & -0.0456(00021) & -0.0545(00044) & 1.1584(00131) & 0.0373(00003) \\
 & $c$ & 0.848 & -0.0466(00058) & -0.0361(00092) & 1.1128(00115) & 0.0309(00003) \\
 & $o$ & 0.879 & -0.0330(00031) & -0.0129(00023) & 1.0717(00111) & 0.0159(00001) \\
\hline
\end{tabular}
\end{table*}
}

\section{New phenomenological model}\label{sec:Model}

First, we phase-folded the time axis of the LCs of each object using the usual equation:
\begin{equation}\label{ec:eq1}
    \phi = \frac{T-T_0}{P}- \text{Int}\left(\frac{T-T_0}{P}\right)\ ,
\end{equation}

\noindent where $\phi$ is the phase, $T$ is the time, $T_0$ is an arbitrary reference time, and $P$ is the period.
We set $T_0$ so that it corresponds to the time of a primary minimum, thus ensuring that phase zero corresponds to the deeper eclipse in the phase-folded LCs. After that, we converted the magnitudes of each filter into normalized fluxes \citep{2016pglp.book.....W} using the following equation:

\begin{equation}\label{ec:eq2}
    I[m(\phi)]=10^{-0.4\times\left[m(\phi)-m(\text{max})\right]}\ ,
\end{equation}

\noindent where $m(\phi)$ is the magnitude at phase $\phi,$ and $m(\text{max})$ corresponds to the maximum value of the magnitude in each filter. We then performed a model based on a so-called super Gaussian \citep[sG;][]{murakami2021pips}, formulated as follows:

\begin{equation}\label{ec:eq3}
    I(\phi)_{\text{sG}}=B_0-B_1\ \text{exp}\left\{-\left[\frac{(\phi-\mu)^2}{2\sigma^2}\right]^p\right\}\ ,
\end{equation}

\noindent where $B_0$ is the mean of the non-varying ground base, $B_1$ is the height, $\mu$ is the mean (location), $\sigma$ is the width, and $p$ is the squareness of the sG. Our model is a combination of this sG and a truncated third-order Fourier fit, given by

\begin{equation}\label{ec:eq4}
    I(\phi)_{\text{Fourier}}=a_0+\sum^3_{n=1}\left[a_n\cos(2\pi n\phi)+b_n\sin(2\pi n\phi)\right]\ ,
\end{equation}

\noindent where $a_0$ corresponds to the mean, and $(a_n,b_n)$ are the $n$-th Fourier coefficients. The modeling process is composed of five stages: 
\begin{enumerate}
    \item First, we calculate $m(\text{max})$  by performing a simple second-order polynomial fit around the brighter maximum. Magnitudes are then converted into normalized flux using Eq.~\ref{ec:eq2}, whereupon we remove bad-quality data by filtering out all points with errors $> 0.08$ in normalized flux.
    \item Following that, we extend the LC for each filter from phases $\phi=-0.2$ to $\phi=1.2$. This is done to facilitate the modeling of the primary eclipses, which take place at phase 0.
    \item We then model the two eclipses independently using the sG from Eq.~\ref{ec:eq3}. The primary eclipse is modeled over the range $\phi=0.8$ to $\phi=1.2$ in phase (hereafter the sG-1 component), whereas for the modeling of the secondary we use phases from $\phi=0.2$ to $\phi=0.8$ instead (the sG-2 component).
    \item Next, we replicate the same sG primary eclipse model at $\phi=1.0$ (i.e., the sG-1 component above) to match the primary eclipse centered on phase $\phi=0$ (the sG-3 component)\footnote{As shown in Fig.~\ref{fig:mod}, sG-1 and sG-3 only differ in terms of their fitted baselines, $B_0$. This happens because we use \texttt{lmfit}’s ``composite modeling'' approach, which combines each user-defined model component (sG-1, sG-2, sG-3, and Fourier, in our case) strictly by summation and fits their respective parameters independently. In other words, \texttt{lmfit} has no built-in way to tie sG-1’s baseline to sG-3’s, so it simply adjusts each of these components' baselines so as to cancel out the local out-of-eclipse flux in their respective phase windows.}.
    \item Finally, when we find the best model for each section of the LC, all of them are combined, and the third-order Fourier fit given by Eq.~\ref{ec:eq4} is added:
    
\begin{equation}\label{ec:eq5}
    I(\phi)_{\text{fit}}= I(\phi)_{\text{sG}*}+I(\phi)_{\text{Fourier}}\ .
\end{equation}

\noindent Here $I(\phi)_{\text{sG}*}$ corresponds to the sum of the three sG models described above, and $I(\phi)_{\text{Fourier}}$ is the truncated third-order Fourier fit, which we use to ensure that the sum of the three sG models properly matches the combined data. Initial parameters for this Fourier fit are estimated using the \texttt{symfit python} library \citep{martin_roelfs_2021_5519611}. Although modeling is done from phases $\phi=-0.2$ to $\phi=1.2$, we use the LC fit from phases 0 to 1 for further analysis. An example of the four components and the final model is presented in Fig.~\ref{fig:mod}.
\end{enumerate}

Appendix~\ref{app:model} describes the difference between our model and the most commonly used 12-term Fourier fit \citep{2009SASS...28..107W}.
The results of the LC fit to our 14 O'Connell EB candidates, which was carried out using \texttt{lmfit} \citep{lmfit2014,lmfit2016} in \texttt{python}, are presented in Appendix~\ref{app:lcs}.

Using the model thus obtained, we next calculate $\Delta I_{\text{max}}$ using Eq.~\ref{ec:eq2}. This is done using two different methods. In the first method, we calculate $\Delta I_{\text{max}}$ from the model values at quadrature phases. This value is denoted in Table~\ref{tab:t2} as $\Delta I_{\text{sG}}$. Uncertainties in this quantity were computed using a bootstrap approach and assuming that the errors of the measurements were normally distributed. 
In the second method, we again perform a simple second-order polynomial fit around each of the maxima derived from our model. This gives us another value for the difference between maxima, which is denoted $\Delta I_{\text{poly}}$ in Table~\ref{tab:t2}. The difference between $\Delta I_{\text{poly}}$ and $\Delta I_{\text{sG}}$ is further explored in Appendix~\ref{app:comp}.

Following the definition provided by \cite{1999PhDT........38M}, we also calculate the OER and the LCA as

\begin{equation}\label{ec:eq6}
    \text{OER}=\frac{\int_{0.0}^{0.5}\left[I(\phi)_{\text{fit}}-I_0\right]d\phi}{\int_{0.5}^{1.0}\left[I(\phi)_{\text{fit}}-I_0\right]d\phi}\
\end{equation}

\noindent and 

\begin{equation}\label{ec:eq7}
    \text{LCA}=\left\{\int_{0.0}^{0.5}\frac{\left[I(\phi)_{\text{fit}}-I(1.0-\phi)_{\text{fit}}\right]^2}{I(\phi)^2_{\text{fit}}}d\phi\right\}^{1/2},
\end{equation}

\noindent respectively, where $I(\phi)_{\text{fit}}$ is our model given by Eq.~\ref{ec:eq5} and $I_0$ is our baseline, defined as the minimum value of our model (at $\phi=0$).

A user-friendly version of the code used for the phenomenological modeling described in this section is available on \texttt{GitHub}\footnote{ 
\url{https://github.com/DavidFloresC/oconnell-modeling}.}.

\section{Results}\label{sec:results}
The results for all 14 objects, including our measured $\Delta I$, OER, and LCA values,  are provided in Table~\ref{tab:t2}. The uncertainties in the derived OER and LCA values were calculated using Monte Carlo integration to solve Eqs.~\ref{ec:eq6} and \ref{ec:eq7} with a 95\% confidence interval. 

\begin{figure}[t]
    \centering
    \includegraphics[width=\columnwidth]{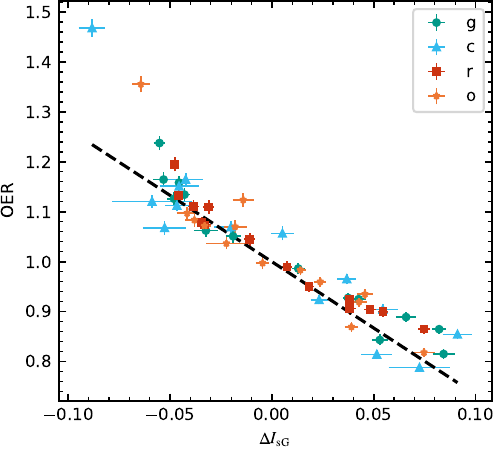}
    \caption{Relation between the OER and $\Delta I_{\text{sG}}$ for the 14 EBs in our sample. Different symbols and colors denote different filters, namely $r$, $g$, $c$, and $o$, following the key at the top right of the plot. The dashed line represents a least-squares fit to the data for all filters combined. The coefficient of determination is $r^2=0.831$, and the resulting fit is given by $\text{OER}=-2.659 \, \Delta I_{\text{sG}} + 1.0$.}
    \label{fig:oer_i}
\end{figure}

\begin{figure}[ht]
    \centering
    \includegraphics[width=\columnwidth]{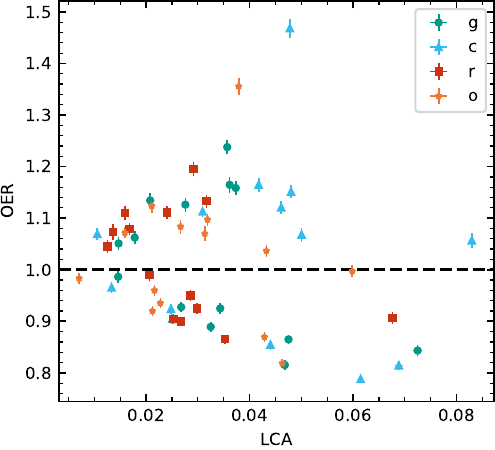}
    \caption{Same as Fig.~\ref{fig:oer_i} but for the relation between the OER and LCA. The dashed line represents the $\text{OER}=1$ locus.}
    \label{fig:lca_oer}
\end{figure}

As expected, a strong anticorrelation between $\Delta I$ and OER can be seen in Fig.~\ref{fig:oer_i}. In Fig.~\ref{fig:lca_oer}, where OER values are shown as a function of LCA, a bifurcated distribution of OER values around $\text{OER}=1$ can be seen, confirming previous findings \citep{1999PhDT........38M}. Stars in our sample are, by and large, split between a subset whose OER values increase with LCA and another for which the OER decreases with increasing LCA, with few exceptions; in both cases, this indicates increasing asymmetry for larger LCA values. Asymmetry tends to be greater in the bluer passbands, again as expected, as cooler spots on the stellar surface have a stronger contrast against their surroundings in the blue due to the increased sensitivity of shorter wavelengths to temperature variations. 

\begin{figure*}[ht]
 \makebox[\textwidth][c]{\includegraphics[width=1.0\textwidth]{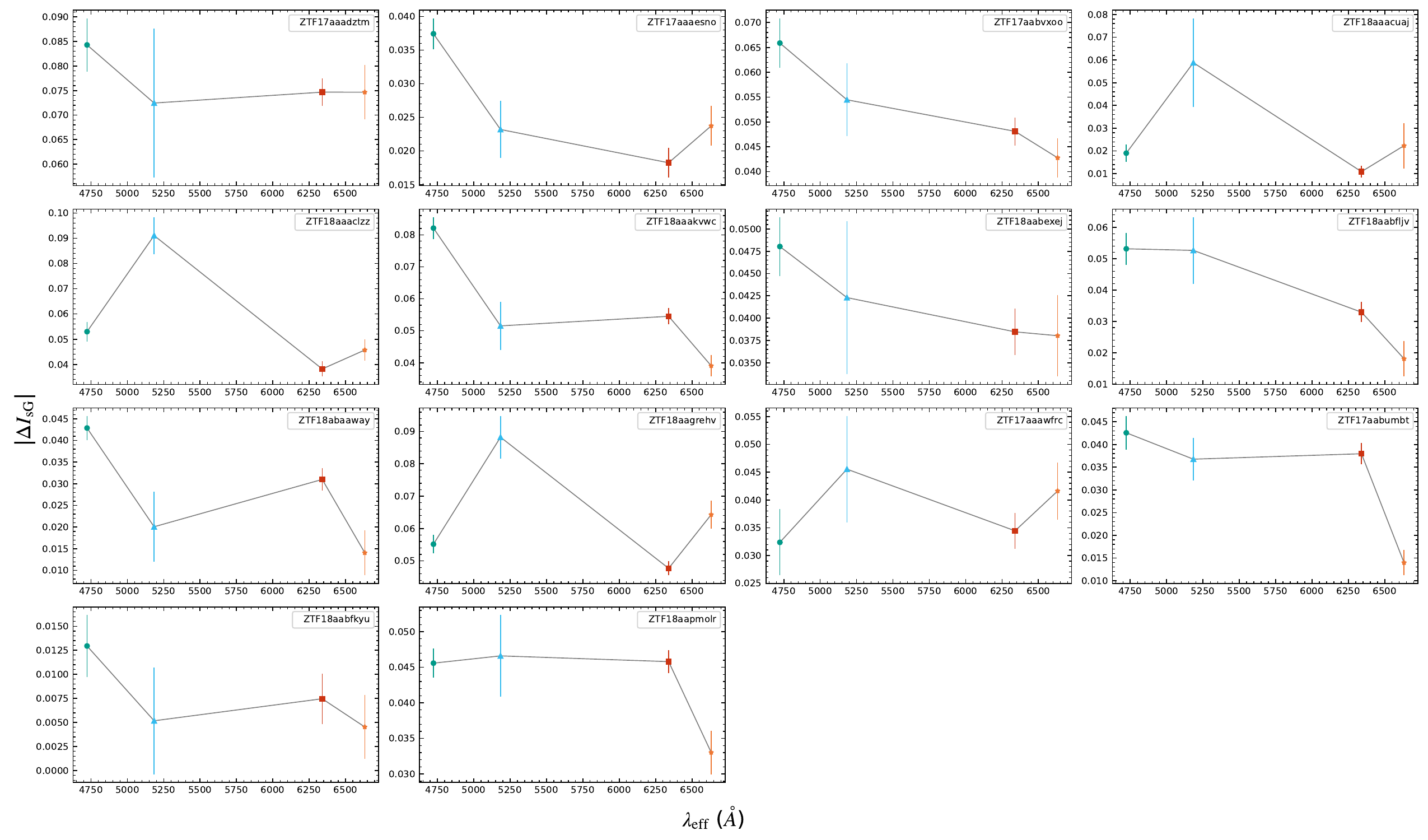}}
 \caption{Absolute values of the difference between the maxima for all the objects in each of the four filters used in this study, plotted as a function of their effective wavelengths. Symbols have the same meaning as in Fig.~\ref{fig:oer_i}.}
 \label{fig:deltas}
\end{figure*}

\begin{figure}
\includegraphics[width=0.475\textwidth]{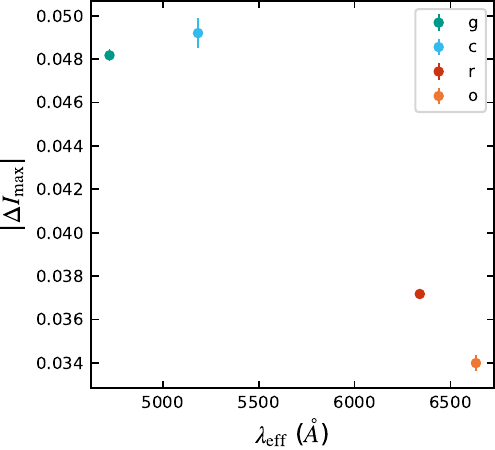}
 \caption{Absolute mean values of the difference between the maxima for all the objects in each of the four filters used in this study, plotted as a function of their effective wavelengths. Symbols have the same meaning as in Fig.~\ref{fig:oer_i}.}
 \label{fig:deltas_mean}
\end{figure}

This is explored further in Fig.~\ref{fig:deltas}, where we examine how the difference between the maxima varies as a function of the effective wavelength ($\lambda_{\text{eff}}$) for each star in our sample. While for most objects the absolute $\Delta I_{\text{sG}}$ value indeed decreases toward the redder filters, there are exceptions~-- which is perhaps not unexpected, since the filters we used are broad and have significant overlap in wavelength coverage (compare the ZTF and ATLAS transmission curves in \citealt{Bellm_2019} and \citealt{2018PASP..130f4505T}, respectively). Figure~\ref{fig:deltas_mean} is similar to Fig.~\ref{fig:deltas} but shows instead mean $\Delta I_{\text{sG}}$ values for each band over the entire sample. There is again a clear trend, with the exceptions of filters $g$ and $c$, whose behavior may also be partly explained by their significant overlap in wavelength coverage.

To explore correlations between $\Delta I_{\text{sG}}$ and other stellar properties, we used Spearman’s correlation coefficient, $\rho$ \citep{10.2307/1412159} and its associated $p$-value. This $p$-value helps indicate if the values of $\rho$ are consistent with the null hypothesis that the characteristics are uncorrelated with $\Delta I_{\text{sG}}$.

\begin{figure*}[ht]
\centering
\includegraphics[width=0.9\textwidth,height=11cm]{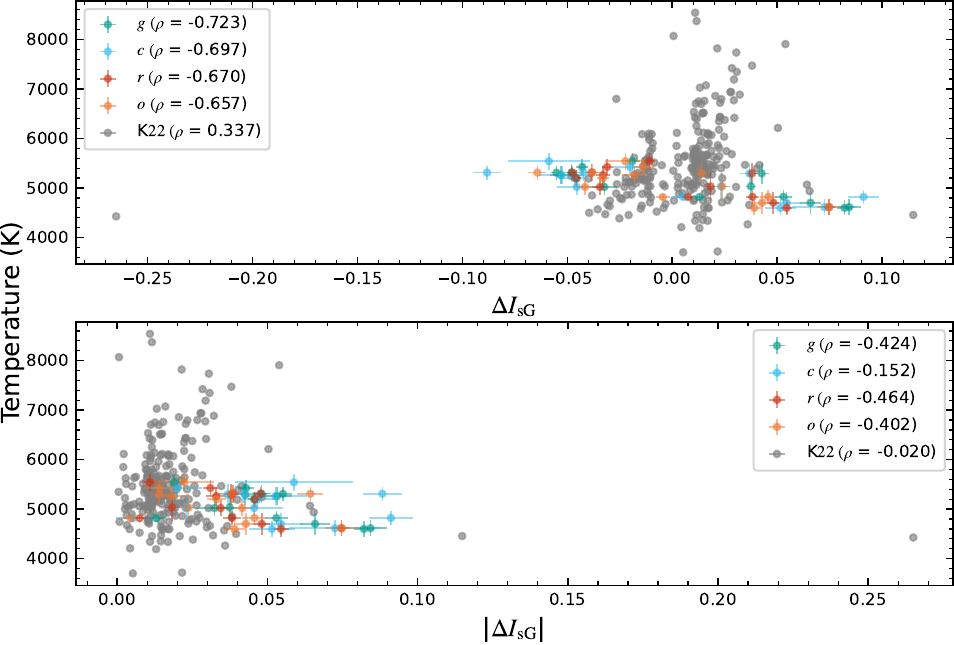}
 \caption{Effective temperatures of our sample stars plotted as a function of $\Delta I_{\rm sG}$ (upper panel) and $|\Delta I_{\rm sG}|$ (bottom panel). Symbols have the same meaning as in Fig.~\ref{fig:oer_i}. Gray corresponds to data from \citet{Knote_2022}.}
 \label{fig:temp_deltas}
\end{figure*}

Unlike \citet{Knote_2022}, we find that $\Delta I_{\text{sG}}$ is significantly correlated with both $T_{\rm eff}$ and the period. The period--color and period--temperature relations in contact EBs that underpin this result have been known for a long time \citep[e.g.,][]{Eggen_1967,Gettel_2006,Qian_2017,Jayasinghe_2020}. In Fig.~\ref{fig:temp_deltas} we present the trends between $\Delta I_{\text{sG}}$ and temperature (upper panel) and $|\Delta I_{\text{sG}}|$ and temperature (bottom panel) alongside Knote's sample. We find that the anticorrelation between $T_{\rm eff}$ and $\Delta I_{\text{sG}}$ in each filter is stronger than the corresponding anticorrelation involving $|\Delta I_{\text{sG}}|$ in the same filter. In particular, we find that a positive O'Connell effect, in which the brighter maximum occurs before the secondary minimum, mainly occurs in relatively cool systems ($T_{\rm eff}\lesssim 5000$~K), while a negative O'Connell effect occurs mainly in systems with $T_{\rm eff}\lesssim 6000$~K. This is supported by \citet{Knote_2022}, who find a similar temperature threshold for negative O'Connell systems. Both of these results agree with the idea that starspots might be the main cause of the O'Connell effect, as their impact is expected to be more pronounced in cooler stars \citep{2009A&ARv..17..251S}. 

\begin{table}[t]
 \caption{Spearman's correlation coefficients ($\rho$) for a positive and negative O'Connell effect ($\Delta I_{\text{sG}}$) and the temperature with their corresponding $p$-values.}
 \label{tab:t3}
 \begin{tabular}{lcc}
  \hline\hline
  Filter & $\Delta I_{\text{sG}}$ & $\Delta I_{\text{sG}}$\\
   & $+$ & $-$ \\
  \hline
  $g$ & -0.714 ($p=0.071$) & 0.214 ($p=0.645$)\\
  $c$ & -0.321 ($p=0.482$) & 0.0 ($p=1.0$)\\
  $r$ & -0.714 ($p=0.071$) & 0.536 ($p=0.215$)\\
  $o$ & -0.600 ($p=0.208$) & -0.047 ($p=0.911$)\\
  \hline
 \end{tabular}
\end{table}

\begin{table}
 \caption{Same as Table~\ref{tab:t3} but using the period instead of the temperature.}
 \label{tab:t4}
 \begin{tabular}{lcc}
  \hline\hline
  Filter & $\Delta I_{\text{sG}}$ & $\Delta I_{\text{sG}}$\\
   & $+$ & $-$ \\
  \hline
  $g$ & -0.143 ($p=0.760$) & -0.536 ($p=0.215$)\\
  $c$ & -0.321 ($p=0.482$) & 0.107 ($p=0.820$)\\
  $r$ & -0.143 ($p=0.760$) & -0.642 ($p=0.119$)\\
  $o$ & -0.600 ($p=0.208$) & -0.405 ($p=0.320$)\\
  \hline
 \end{tabular}
\end{table}

Although the anticorrelation between $\Delta I_{\text{sG}}$ and temperature is similar for different filters, it becomes slowly less pronounced toward redder effective wavelengths. When comparing positive and negative O'Connell effects (as revealed by the $\Delta I_{\text{sG}}$ values), we observe differences in correlations between  $\Delta I_{\text{sG}}$  and the aforementioned parameters (see Table~\ref{tab:t3}); despite the lower statistical significance due to the reduced sample sizes, in almost all filters the correlation has different signs, which depend on the sign of the O'Connell effect. The exception to this is the $o$ filter, which shows the same anticorrelation for both positive and negative $\Delta I_{\text{sG}}$. Another particular case is $c$, which shows no correlation at all for negative $\Delta I_{\text{sG}}$. While a much larger sample is needed to put these results on a firm footing, this preliminary analysis supports the idea that the O'Connell effect is caused by different mechanisms, depending on the $\Delta I_{\text{sG}}$ sign. Unlike \citet{Knote_2022}, we find that $|\Delta I_{\text{sG}}|$ does not present significant correlations, and also that $\Delta I_{\text{sG}}$ has a strong anticorrelation with the temperature. Although statistical uncertainties are much larger in our case, our sample seems to behave differently than theirs. This difference might be explained by the fact that their sample lacks systems with larger $\Delta I_{\text{sG}}$ values. Figure~\ref{fig:temp_deltas} shows how the majority of our sample extends beyond both extremes of the data in their study, and how the correlation is much weaker, especially for positive values of $\Delta I_{\text{sG}}$. These results are also consistent when we limit Knote's sample to between $4500$ and $6000$~K to match temperatures similar to those of our sample. When this limit is imposed on Knote's sample, we find similar correlation coefficients: ($\rho=0.269$, $p=0.0003$) and ($\rho=-0.135$, $p=0.075$) for $\Delta I_{\text{sG}}$ and $|\Delta I_{\text{sG}}|$, respectively. This might suggest that systems with higher $\Delta I_{\text{sG}}$ values (negative or positive) do not behave the same way as systems with a weaker O'Connell effect. To further investigate this possibility, a large sample of systems with high $\Delta I_{\text{sG}}$ values and a much wider range of temperatures than in our study is needed.

\begin{figure*}
\centering
 \includegraphics[width=0.9\textwidth,height=11cm]{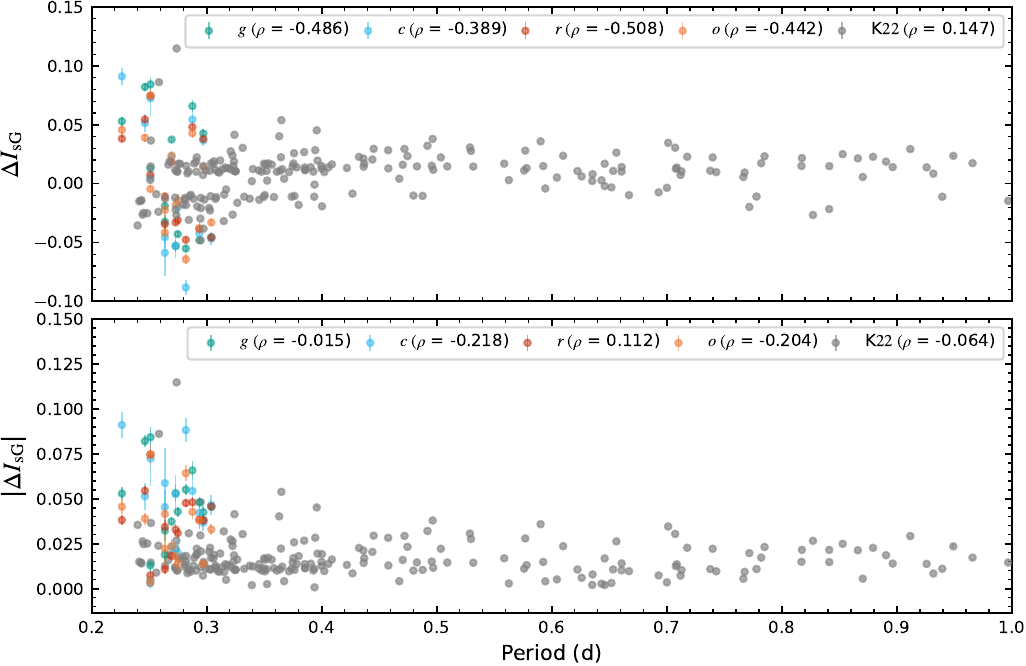}
 \caption{Correlation between $\Delta I_{\rm sG}$ and the period for our sample. Symbols have the same meaning as in Fig.~\ref{fig:oer_i}. Gray corresponds to data from \citet{Knote_2022}, limited to the range 0.2 to 1 days for clarity.}
 \label{fig:period_deltas}
\end{figure*}

The other correlation worth noticing is the one between $\Delta I_{\text{sG}}$ or $|\Delta I_{\text{sG}}|$ and the period (Fig.~\ref{fig:period_deltas}, upper and lower panels, respectively). Similarly to correlations involving the temperature, we find a stronger anticorrelation with $\Delta I_{\text{sG}}$ than $|\Delta I_{\text{sG}}|$ across all passbands. The correlations are shown in Table~\ref{tab:t4}, separately for positive and negative $|\Delta I_{\text{sG}}|$ systems. We find that generally this anticorrelation is stronger when the O'Connell effect has a negative sign. Although this is not in agreement with the results obtained by \citet{Knote_2022} --- who find that a correlation between $\Delta I_{\text{sG}}$ and the period (the ``O'Connell effect size'' in their terminology) is slightly positive in all cases, but stronger in negative O'Connell effect systems --- they also find that these correlations are noticeable for systems with $P\leqslant 0.5$~d. This is also true for our sample with short periods, $P\leqslant 0.3$~d.

The presence of correlations for $\Delta I_{\text{sG}}$ but not $|\Delta I_{\text{sG}}|$ indicates that the physical mechanisms driving positive and negative O'Connell effects may differ fundamentally. While starspots have been proposed as the dominant explanation, with larger $\Delta I_{\text{sG}}$ values corresponding to greater spot coverage \citep[][]{10.1093/pasj/psy140,Knote_2022}, our results reveal some inconsistencies with this interpretation.

The observed anticorrelations with both period and temperature are consistent with the known period--color relation in contact binaries, where shorter-period systems tend to be cooler. However, if spots were the primary cause, we would expect $|\Delta I_{\text{sG}}|$ to correlate strongly with both parameters, as demonstrated by \citet{10.1093/pasj/psy140} for W-type (W Ursae Majoris, or EW) binaries (see Sect.~\ref{sec:conclusions}). Instead, we find these correlations to be weak or absent in certain passbands.

The band dependence of these trends, including the complete lack of correlation for negative $\Delta I_{\text{sG}}$ in the $c$ filter, further complicates the spot interpretation. While our results remain qualitatively consistent with \citeauthor{10.1093/pasj/psy140} (\citeyear{10.1093/pasj/psy140})'s findings regarding spot properties, the quantitative discrepancies suggest that additional mechanisms may contribute to the O'Connell effect, particularly for systems exhibiting large asymmetries. This conclusion aligns with recent suggestions that close binary interactions may play a significant role in generating the observed LC asymmetries \citep{GUO2025102341,Wang_2025}.

\section{Discussion and conclusions}\label{sec:conclusions}
As shown in Table~\ref{tab:t2}, the percentage difference between maxima is in the range [$1\%$-$13\%$], while the ratio of positive-to-negative signs of $\Delta I_{\rm sG}$ is close to 0.5. Cross-matching our sample with \textit{TESS} Input Catalog Version 8 \citep{2019AJ....158..138S}, we found that all of the 14 EBs in our sample are late-type contact binaries with spectral types [K4V-G7V] that have effective temperatures ($T_{\rm eff}$) cooler than 5600~K, as shown in Table~\ref{tab:ztflist}. Combined with the orbital periods ($\leq 0.3$~d), this suggests that the contact EBs are rather W subtypes. Since both components are fast-rotating cool stars, their surfaces should show deep convective activity. Thus, the most probable mechanism for the LC asymmetries in our sample is related to the presence of magnetic fields. In addition, due to high chromospheric activity, one might expect these systems to also show strong X-ray emissions. Nevertheless, none of these EBs are found in the fourth XMM (X-ray Multi-Mirror Mission)-\textit{Newton} serendipitous source catalogue \citep{2020A&A...641A.136W} within a $10''$ search radius. The absence of X-ray detections is most likely attributable to the faintness of these particular systems. This is unlike brighter late-type contact binaries, which are frequently identified as X-ray emitters and for which it is often feasible to model their X-ray spectra \citep[e.g.,][]{2001ApJ...562L..75B}. Despite the X-ray non-detections, data from the Galaxy Evolution Explorer (GALEX) Data Release 6 \citep[][]{2014AdSpR..53..900B} indicate that 6 out of these 14 EBs (within a $3''$ radius) are likely associated with near-UV emission, which can also be an indicator of chromospheric activity.

Spectroscopic observations around the activity-sensitive Ca\,{\sc II}~H\&K and Balmer (H$\alpha$) spectral lines \citep{1966ApJ...144..695W, 2021ApJS..253...51L} could show equivalent width variations and/or emission due to chromospheric activity. Long-term photometric monitoring could also reveal out-of-eclipse light variations as the result of starspot migration and/or evolution or the presence of magnetic activity cycles \citep{2017MNRAS.465.4678M, 2018ApJS..238....4P, 2020AJ....160...62H}. Radial velocities for the brighter systems, combined with archival multiband photometric LCs, will provide accurate absolute parameters for the components of the systems. For the total eclipsing contact system candidates (e.g., ZTF18aabexej and ZTF18aabfljv), approximate absolute physical parameters could be derived from multiband photometry only \citep{2023AJ....165...80P}. Using their absolute physical parameters, it is possible to evaluate the ability of the stellar engine to produce period and LC variations through the Applegate mechanism \citep{1992ApJ...385..621A}.

We find that, although none of our systems are particularly hot, systems with a negative O'Connell effect ($\Delta I_{\rm sG} < 0$) have higher temperatures in general. This could mean that in these systems there is a correlation between the O'Connell effect and spots. Systems with negative $\Delta I_{\rm sG}$ also have temperatures $T < 6000$ K \citep{Knote_2022}. None of our systems have temperatures greater than $6000$~K, likely due to the small sample size in our study.

These results also show that $\Delta I_{\rm sG}$ is anticorrelated with temperature and period, and that this anticorrelation is stronger when compared with $|\Delta I_{\rm sG}|$. These correlations were found in all the different passbands, and they are consistent for most of them, with the exception of $r$ and its corresponding correlation between $|\Delta I_{\rm sG}|$ and the period, which is different from all the anticorrelations from other passbands. This might be due to the small sample size or the fact that positive and negative O'Connell effects follow different correlations, as seen in Table~\ref{tab:t4}, where we see a significant difference when comparing negative and positive $\Delta I_{\rm sG}$ in the $r$ band. The same seems to happen in $g$, as the corresponding correlation with $|\Delta I_{\rm sG}|$ is nonexistent. This, combined with the fact that $c$ and $o$ follow only weak anticorrelations, suggests that more data are needed to put our results on a firmer footing. These few differences between correlations for positive and negative $\Delta I_{\rm sG}$ seem to indicate that the results from this study are not fully consistent with \citeauthor{10.1093/pasj/psy140} (\citeyear{10.1093/pasj/psy140})'s study on starspots.

In all cases, correlations were much more significant for $\Delta I_{\rm sG}$  than for $|\Delta I_{\rm sG}|$, even though the correlation with the period seems to be weaker in passbands $o$ and $c$. Although the difference is not very large, more data in these two passbands would be useful to obtain more conclusive results. This is also not related to the anticorrelation between $\Delta I_{\rm sG}$ and the temperature, where all passbands correlate relatively similarly with no exceptions, albeit these correlations do seem to become weaker toward longer wavelengths. Observations of a larger sample, particularly at short wavelengths, would certainly be of help in establishing these correlations.

Furthermore, while our targets are intrinsically faint, some of the scatter in the OER, $\Delta I_{\mathrm{sG}}$, and LCA parameters may be due to time variability. The evolution and migration of starspots over time can contribute to this scatter, as evidenced by the dispersion seen in our phase-folded LCs. When the data for our stars were split into two or more segments covering smaller time spans to check for long-term patterns, in almost all cases the LC segments were incomplete, making further analysis unreliable. Thus, although our current dataset covers many cycles, a more detailed temporal analysis with higher-cadence data would help us disentangle intrinsic long-term variability patterns from asymmetry effects that are (nearly) invariant in time.

From this, at least for the stars in our sample, we cannot rely solely on the idea of starspots as the only cause of the O'Connell effect, suggesting that multiple causes may be at play. This is supported by the fact that we observe differences in the correlations between positive and negative values of $\Delta I_{\rm sG}$ and the period and temperature. Although the results for different passbands are not fully conclusive, we also find that, for the majority of our sample, the O'Connell effect is most predominant at short wavelengths. Cool starspots have a stronger contrast against the stellar photosphere in bluer wavelengths, which may contribute to the pronounced effect observed in these bands. Still, starspots might not provide a sufficient explanation for all O’Connell effect systems. A larger sample, and an extension of our coverage toward shorter effective wavelengths, would help in further investigating these results.

Starspots are typically the reason given for the O'Connell effect. While this could be true for some systems, we find evidence that, at least for the stars in our sample, starspots may not be its only cause. If so, the O'Connell effect may also be connected to the interaction between the components rather than, or in addition to, starspots. Extending the size of the samples studied using a multiwavelength approach would be of crucial importance in unveiling the root cause(s) of this intriguing phenomenon.

\begin{acknowledgements}
    We thank an anonymous referee for several useful comments and suggestions.
    Support for this project is provided by ANID's FONDECYT Regular grants \#1171273  and 1231637; ANID's Millennium Science Initiative through grants ICN12\textunderscore 009 and AIM23-0001, awarded to the Millennium Institute of Astrophysics (MAS); and ANID's Basal project FB210003.  AP gratefully acknowledges the support provided by the grant co-financed by Greece and the European Union (European Social Fund~-- ESF) through the Operational Programme ``Human Resources Development, Education and Lifelong Learning'' in the context of the project ``Reinforcement of Postdoctoral Researchers~-- 2nd Cycle'' (MIS-5033021), implemented by the State Scholarships Foundation (IKY). 
    This work has made use of data from the Asteroid Terrestrial-impact Last Alert System (ATLAS) project. ATLAS is primarily funded to search
      for near earth asteroids through NASA grants NN12AR55G, 80NSSC18K0284,
      and 80NSSC18K1575; byproducts of the NEO search include images and
      catalogues from the survey area.  The ATLAS science products have been
      made possible through the contributions of the University of Hawaii
      Institute for Astronomy, the Queen's University Belfast, the Space
      Telescope Science Institute, the South African Astronomical Observatory (SAAO),
      and the Millennium Institute of Astrophysics (MAS), Chile.
\end{acknowledgements}

\bibliographystyle{aa}
\bibliography{biblio}

\begin{appendix}
\clearpage
\section{Model comparison}\label{app:model}

\begin{figure}[hb]
\centering
 \includegraphics[width=0.582\columnwidth]{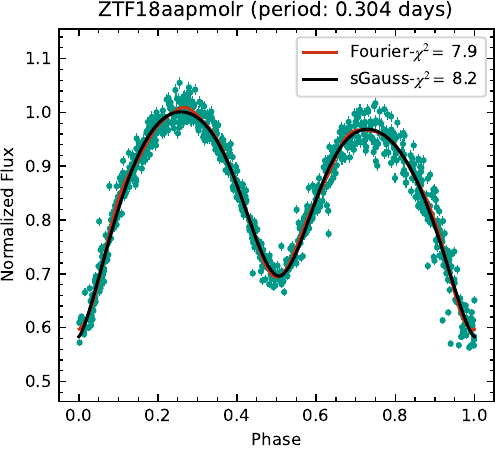}
 \caption{Comparison between a 12-term Fourier fit (red) and the model proposed in this work (black), using ZTF $g$-band data. The reduced $\chi^2$ values for the Fourier and new models are 7.9 and 8.2, respectively.}
 \label{fig:model1}
\end{figure}

\begin{figure}[hb]
\centering
 \includegraphics[width=0.582\columnwidth]{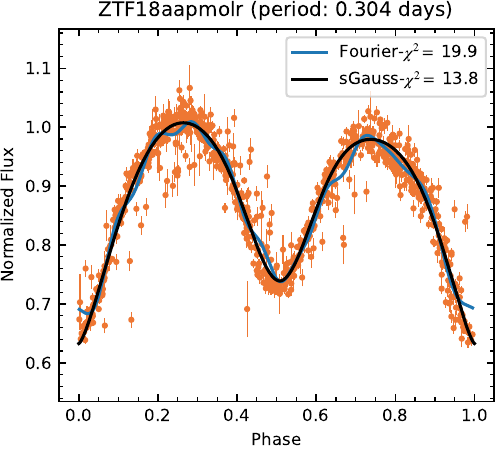}
 \caption{Same as Fig.~\ref{fig:model1} but for the $o$ band. The Fourier fit is shown in blue. The reduced $\chi^2$ values are 19.9 and 13.8, respectively.}
 \label{fig:model2}
\end{figure}

Here we provide a brief comparison between our model and the 12-term Fourier fit proposed by \cite{2009SASS...28..107W}.
The latter method can perform equally or better than the one proposed in this study when well-sampled data are available, as in the case of our ZTF LCs (see Fig.~\ref{fig:model1}). However, in the case of noisier data, as happens with our ATLAS LCs, the 12-term Fourier fit may introduce artifacts, as shown in Fig.~\ref{fig:model2}, whereas our model remains well behaved.\footnote{The \texttt{python symfit} library was used to perform the 12-term Fourier fits.} In most cases, the primary problem is to get a good fit around $\phi=0.5$. While this may not be relevant when measuring $\Delta I_{\rm max}$, it is crucial to calculate accurate OER and LCA values.
The model presented in this work tries to work around that problem by using different sG fits around the different eclipses ($\theta=0,0.5,1$), which ensures that the minima of each LC will be accurately reproduced. In our case, the Fourier fit is then simply used to correct the sum of all three different models and get accurate maxima.

Although our model is more flexible and can be used as an alternative phenomenological description of the LCs of EBs presenting the O'Connell effect, one of the advantages of using a 12-term Fourier fit is that it allows for an early classification of the system using its cosine coefficients, specifically the second and fourth terms, $a_2$ and $a_4$ \citep{Rucinski1997}. This can be of great utility for unclassified systems. We defer finding such correlations with all different parameters that enter our phenomenological model to future work, in which we plan to use an enlarged sample containing previously classified O'Connell effect EBs.

\section{Comparison of the maxima}\label{app:comp}
\begin{figure}[h]
 \centering
 \includegraphics[width=0.475\textwidth]{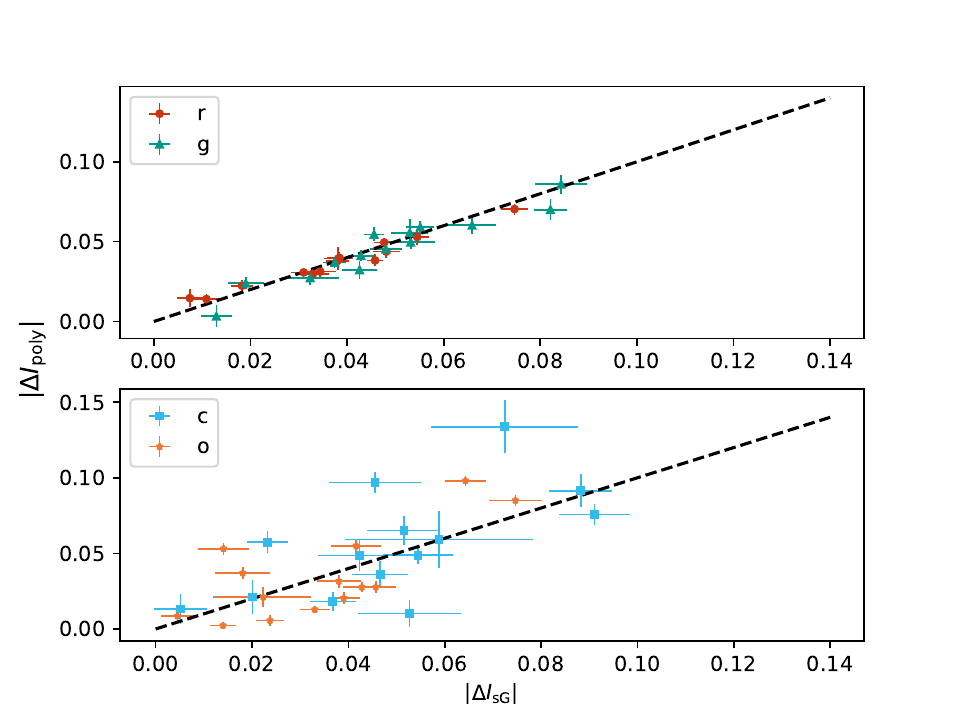}
 \caption{Comparison of |$\Delta I_{\rm max}$|. Top panel: ZTF $g$ and $r$ filters. Bottom panel: ATLAS $c$ and $o$ filters. In both cases the dashed line corresponds to the identity line.}
 \label{fig:maxima_comp}
\end{figure}

\begin{figure}[h]
 \centering
 \includegraphics[width=0.475\textwidth]{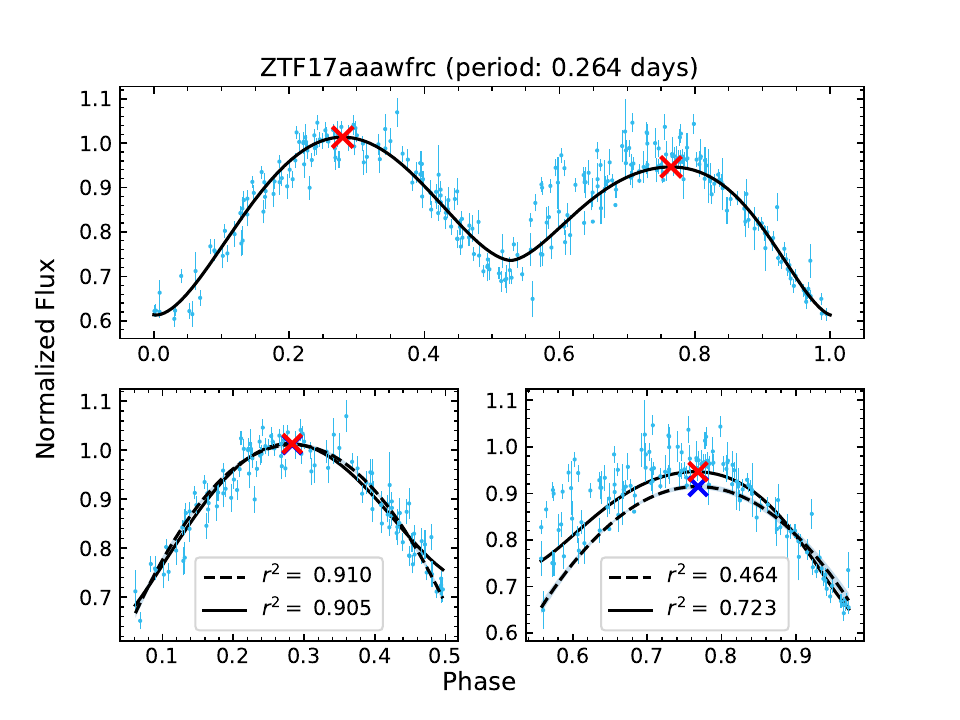}
 \caption{Comparison between the polynomial fit and our proposed model. Top panel: Complete LC of ZTF17aaawfrc in ATLAS $c$. Bottom-left panel: Primary maxima of the LC. Bottom-right panel: Secondary maxima of the LC. The solid black line corresponds to the sG model, and the dashed line to the polynomial fit. Blue and red crosses indicate the maxima obtained using the polynomial fit and the sG model, respectively.}
 \label{fig:maxima}
\end{figure}

Although our model performs better in some cases when calculating O'Connell LC features like OER and LCA, this is not necessarily true for $\Delta I_{\rm max}$. For this reason, we carried out a direct comparison between the two methods used in this work, namely the polynomial fit and our model ($\Delta I_{\rm poly}$ and $\Delta I_{\rm sG}$, respectively, both of which are shown in Table~\ref{tab:t2}). As seen in Fig.~\ref{fig:maxima_comp}, for the ZTF filters the values are sufficiently similar that we cannot distinguish between their results. However, when ATLAS data are used, we do see some important discrepancies in a number of cases; an example of this is shown in Fig.~\ref{fig:maxima}, where we can see a noticeable difference in the inferred secondary maxima.
Through visual inspection, we conclude that the maxima obtained using the sG model (red crosses in Fig.~\ref{fig:maxima}) are more reliable in the majority of cases since.


\onecolumn 
\section{Modeled light curves}\label{app:lcs}

Here we present all the LCs used in this work with each corresponding model described in Sect.~\ref{sec:Model} for all the different passbands.

\begin{figure}[H]
\centering
 \includegraphics[width=0.7\textwidth]{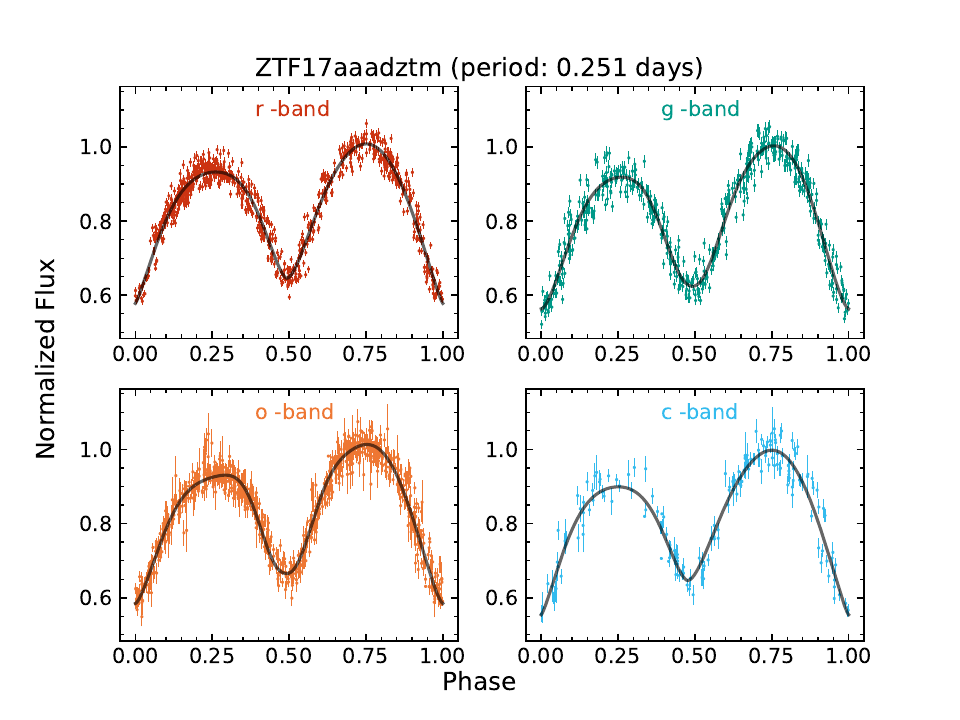}
 \caption{LCs of the O'Connell EB candidate ZTF17aaadztm in the four filters used in this study: ZTF $r$, ZTF $g$, ATLAS $c$, and ATLAS $o$. The gray line corresponds to the best fit using our model (Eq.~\ref{ec:eq5}). The period of the star is also shown at the top of the figure, along with its ZTF ID.\\[12pt]}
 \label{fig:lc1}
 \includegraphics[width=0.7\textwidth]{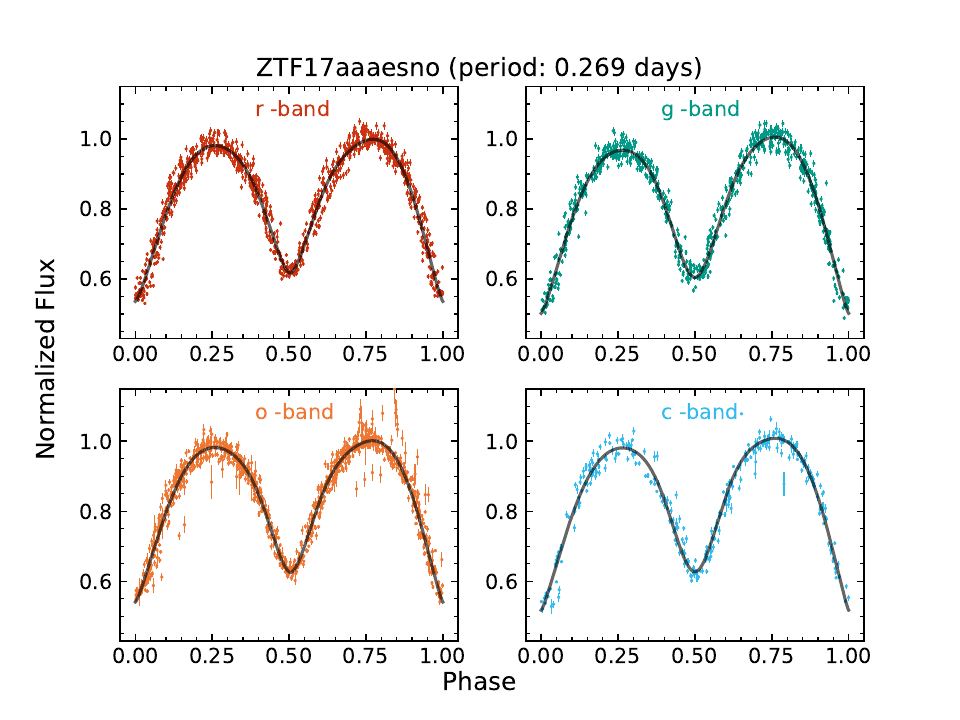}
 \caption{Same as Fig.~\ref{fig:lc1} but for ZTF17aaaesno.}
 \label{fig:lc2}
\end{figure}

\vspace{2cm}

\begin{figure}
\centering
 \includegraphics[width=0.8\textwidth,height=10.5cm]{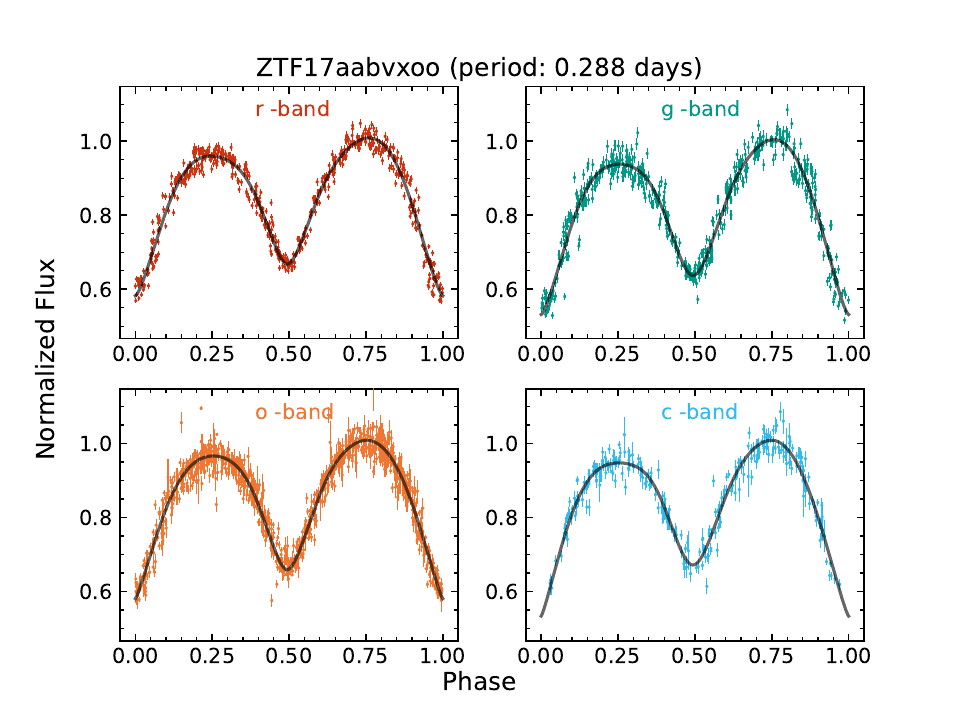}
 \caption{Same as Fig.~\ref{fig:lc1} but for ZTF17aabvxoo.}
 \label{fig:lc3}
\end{figure}
\begin{figure}
\centering
 \includegraphics[width=0.8\textwidth,height=10.5cm]{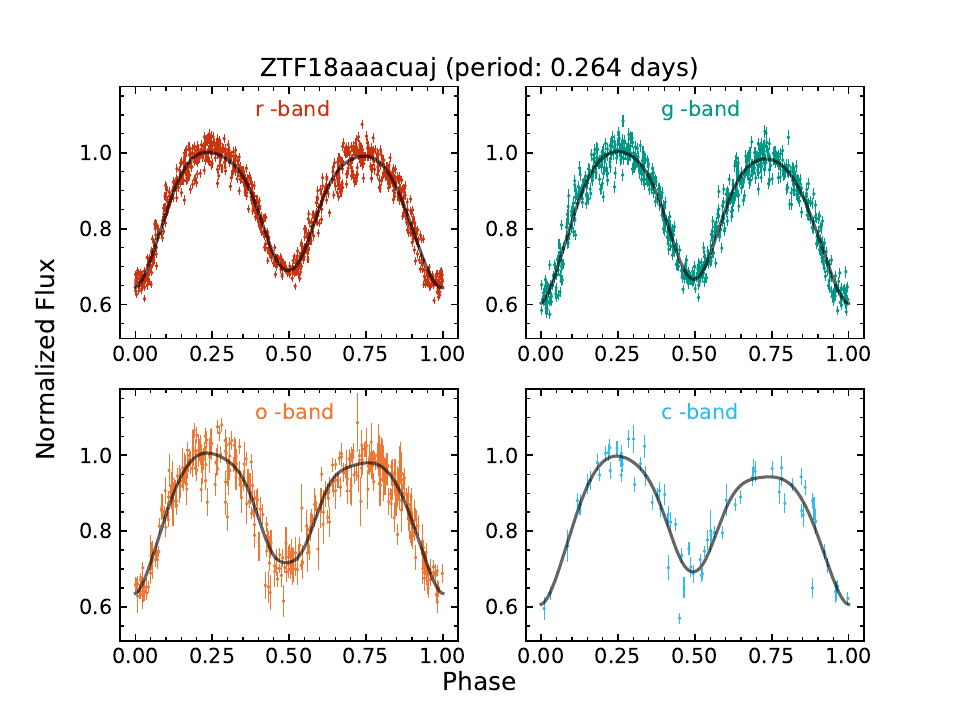}
 \caption{Same as Fig.~\ref{fig:lc1} but for ZTF18aaacuaj.}
 \label{fig:lc4}
\end{figure}
\begin{figure}
\centering
 \includegraphics[width=0.8\textwidth,height=10.5cm]{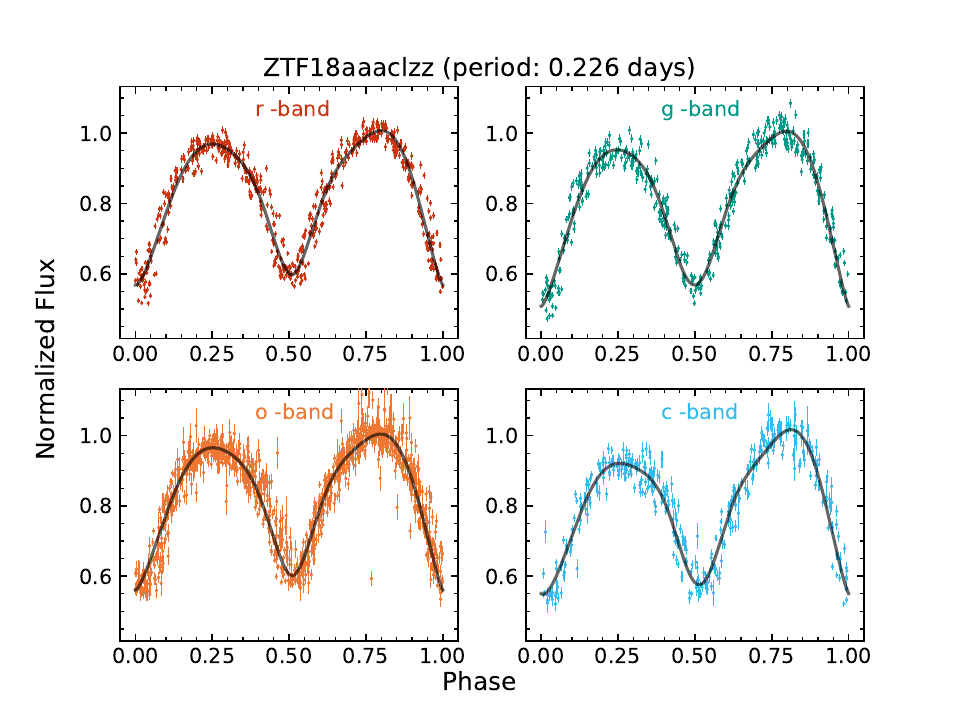}
 \caption{Same as Fig.~\ref{fig:lc1} but for ZTF18aaaclzz.}
 \label{fig:lc5}
\end{figure}
\begin{figure}
\centering
 \includegraphics[width=0.8\textwidth,height=10.5cm]{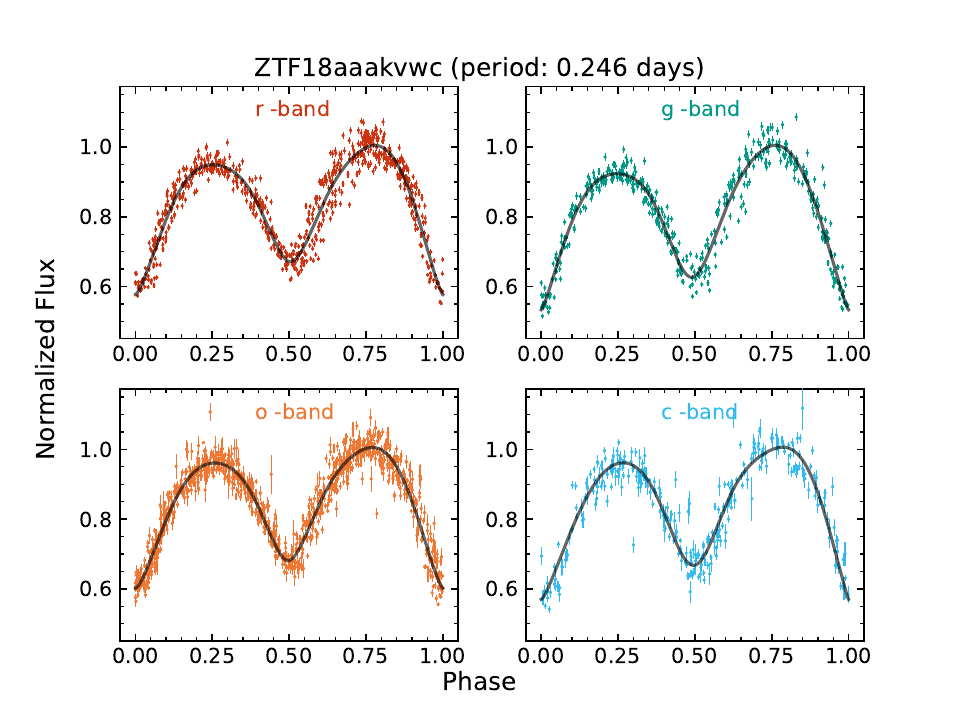}
 \caption{Same as Fig.~\ref{fig:lc1} but for ZTF18aaakvwc.}
 \label{fig:lc6}
\end{figure}
\begin{figure}
\centering
 \includegraphics[width=0.8\textwidth,height=10.5cm]{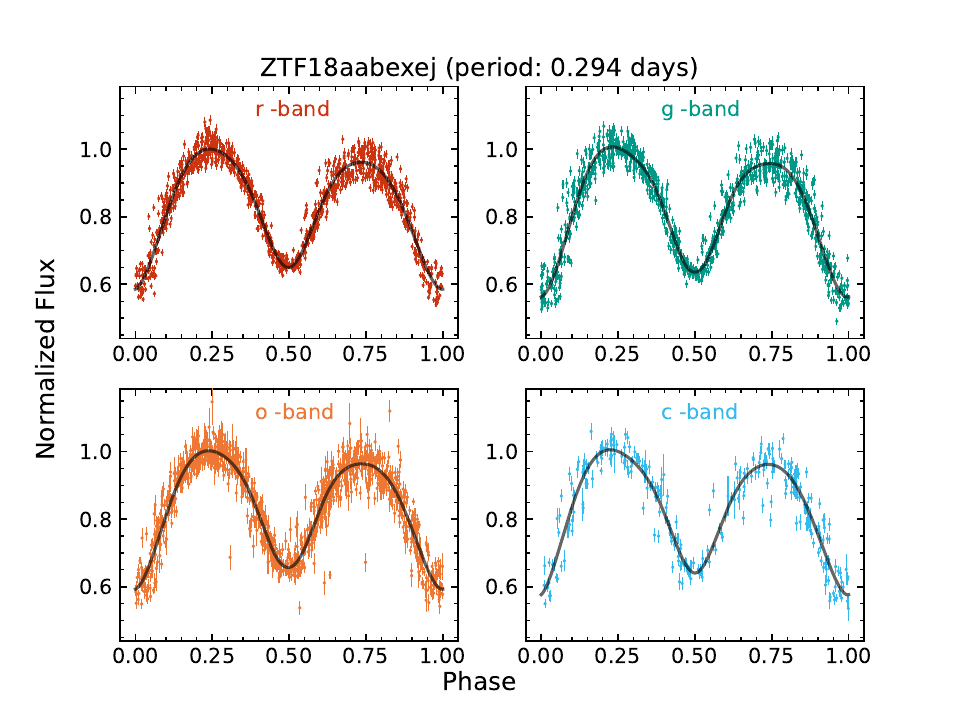}
 \caption{Same as Fig.~\ref{fig:lc1} but for ZTF18aabexej.}
 \label{fig:lc7}
\end{figure}
\begin{figure}
\centering
 \includegraphics[width=0.8\textwidth,height=10.5cm]{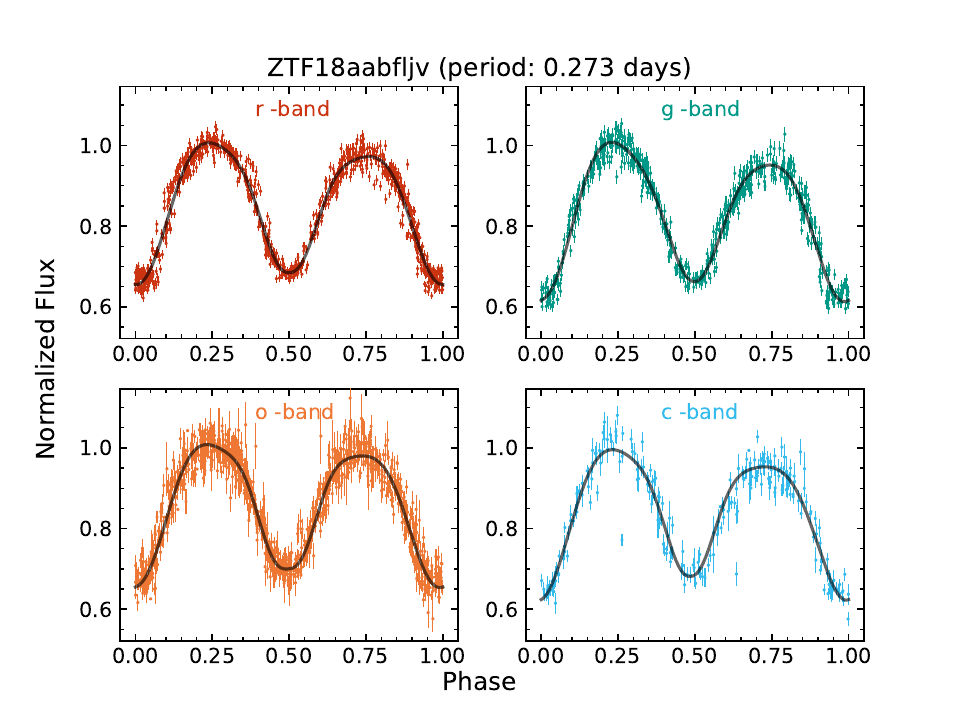}
 \caption{Same as Fig.~\ref{fig:lc1} but for ZTF18aabfljv.}
 \label{fig:lc8}
\end{figure}
\begin{figure}
\centering
 \includegraphics[width=0.8\textwidth,height=10.5cm]{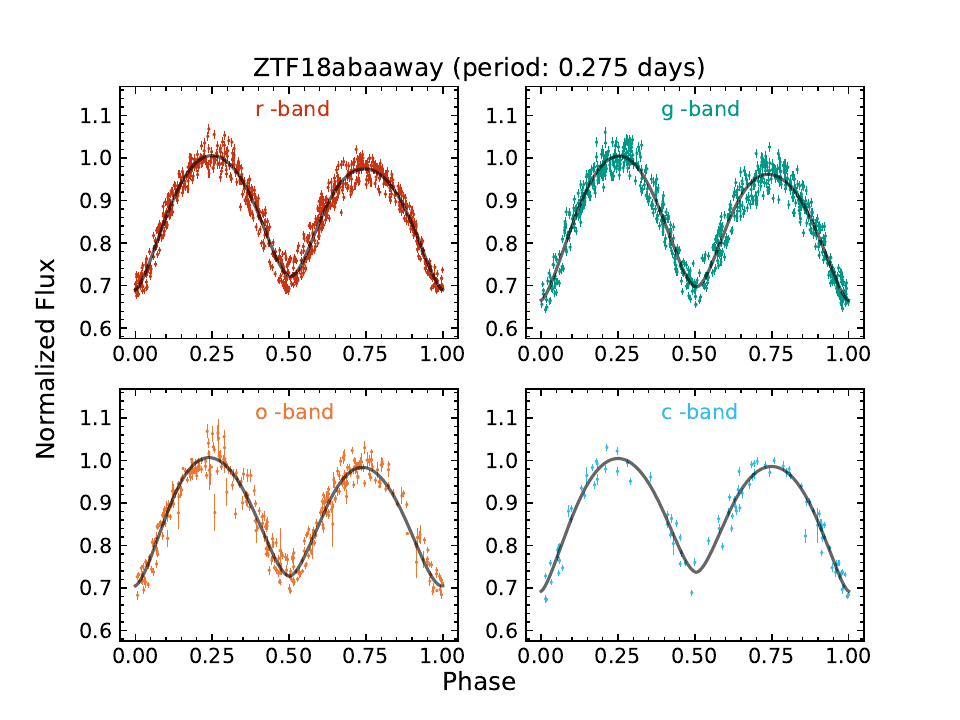}
 \caption{Same as Fig.~\ref{fig:lc1} but for ZTF18abaaway.}
 \label{fig:lc9}
\end{figure}
\begin{figure}
\centering
 \includegraphics[width=0.8\textwidth,height=10.5cm]{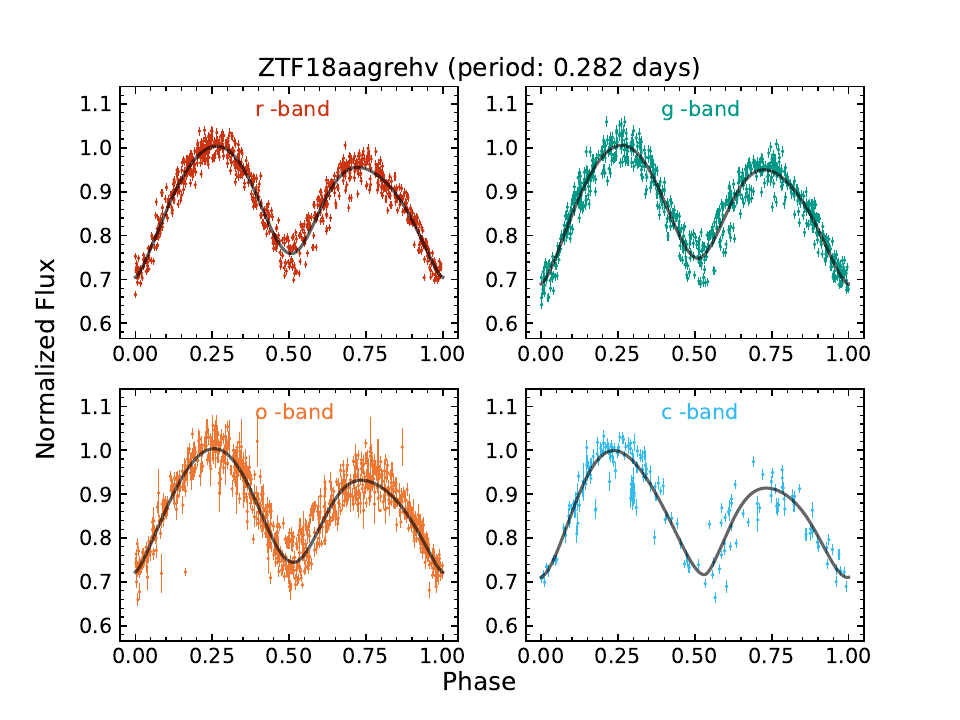}
 \caption{Same as Fig.~\ref{fig:lc1} but for ZTF18aagrehv.}
 \label{fig:lc10}
\end{figure}
\begin{figure}
\centering
 \includegraphics[width=0.8\textwidth,height=10.5cm]{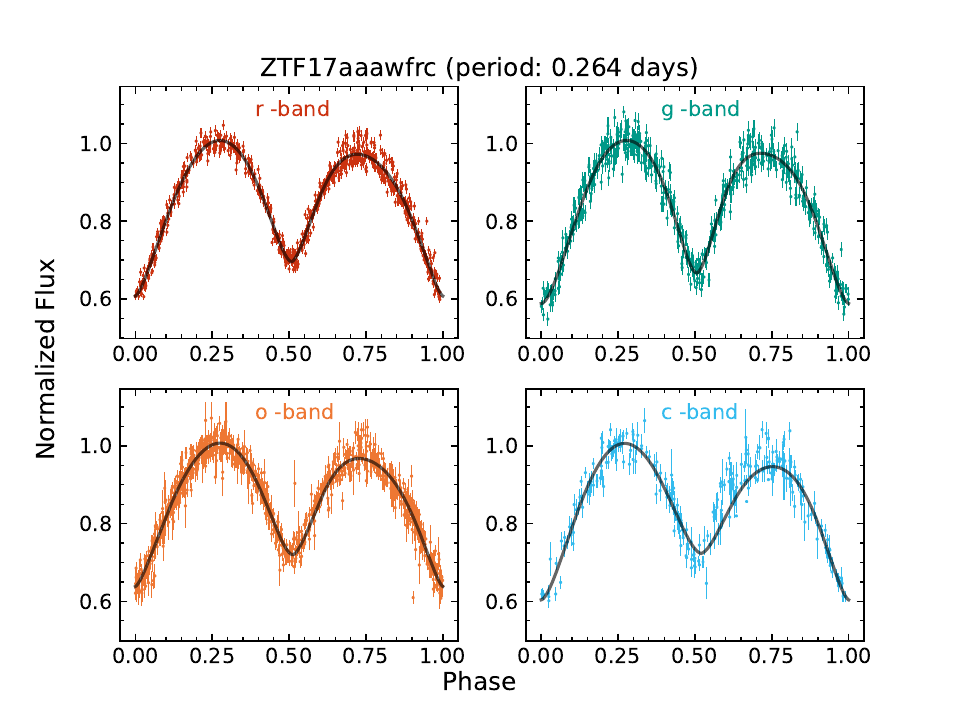}
 \caption{Same as Fig.~\ref{fig:lc1} but for ZTF17aaawfrc.}
 \label{fig:lc11}
\end{figure}
\begin{figure}
\centering
 \includegraphics[width=0.8\textwidth,height=10.5cm]{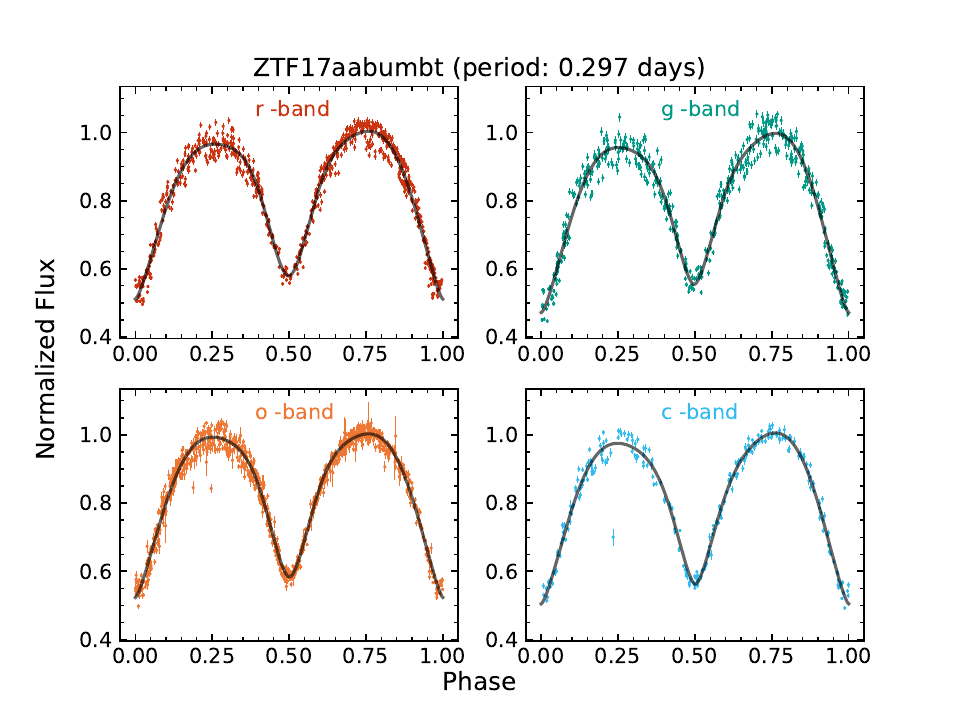}
 \caption{Same as Fig.~\ref{fig:lc1} but for ZTF17aabumbt.}
 \label{fig:lc12}
\end{figure}
\begin{figure}
\centering
 \includegraphics[width=0.8\textwidth,height=10.5cm]{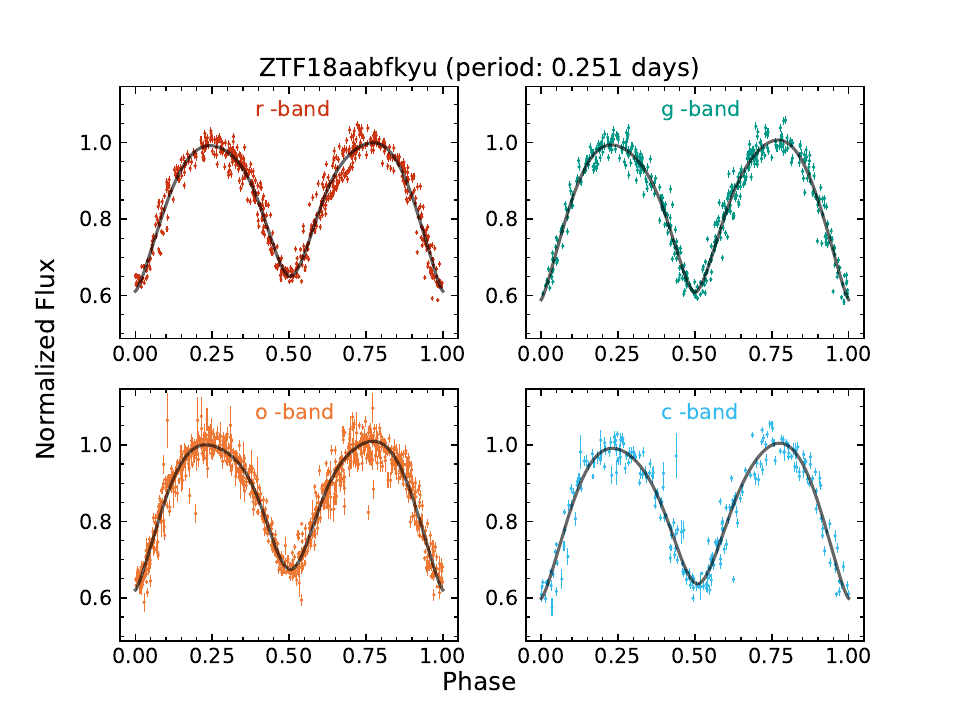}
 \caption{Same as Fig.~\ref{fig:lc1} but for ZTF18aabfkyu.}
 \label{fig:lc13}
\end{figure}
\begin{figure}
\centering
 \includegraphics[width=0.8\textwidth,height=10.5cm]{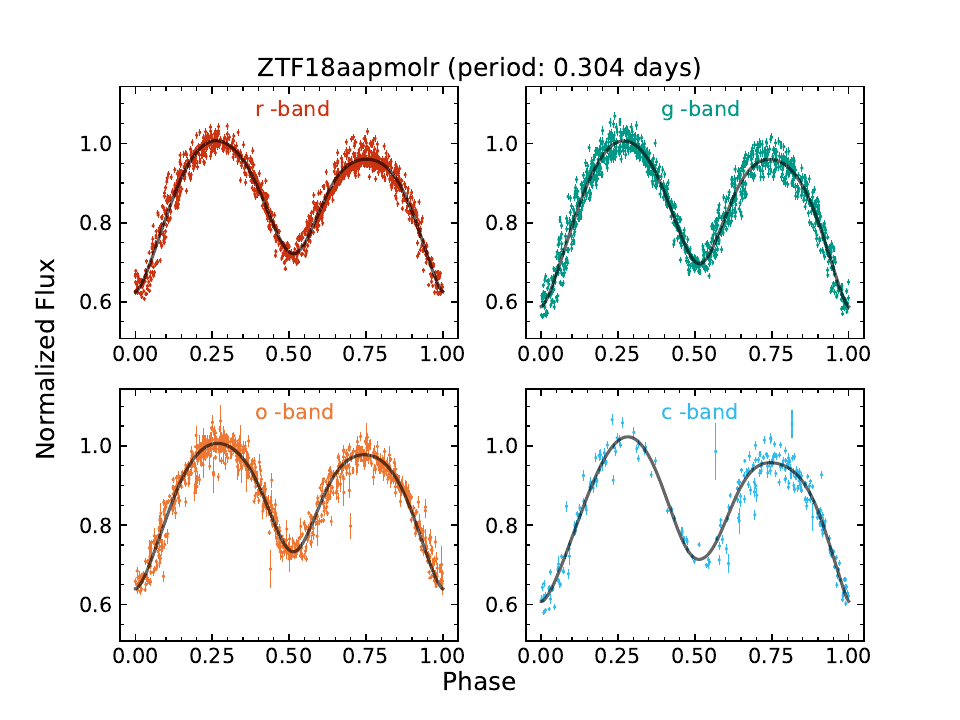}
 \caption{Same as Fig.~\ref{fig:lc1} but for ZTF18aapmolr.}
 \label{fig:lc14}
\end{figure}

\end{appendix}
\end{document}